\documentclass[
superscriptaddress,
nofootinbib,
 twocolumn,
 amsmath,amssymb,
 aps,
]{revtex4-1}

\usepackage[utf8]{inputenc}
\usepackage{graphicx}
\usepackage{xcolor}
\usepackage{hyperref}
\usepackage{siunitx}

\newcommand{\lmn}{{\ell m n}}

\newcommand{\signalinj}{Signal}
\newcommand{\controlinj}{Control}
\newcommand{\bayesfac}{\ensuremath{\mathcal{B}}} 
\newcommand{\eventbftime}{\ensuremath{t_{\rm ref} + 6\,\mathrm{ms}}} 
\newcommand{\eventbf}{{\ensuremath{56 \pm 1 }}} 
\newcommand{\approxebf}{\ensuremath{56}} 

\usepackage{array}
\newcolumntype{P}[1]{>{\centering\arraybackslash}p{#1}}

\begin{document}

\title{Spectroscopy for asymmetric binary black hole mergers}

\author{Jahed Abedi}
\email{jahed.abedi@uis.no}
\affiliation{Department of Mathematics and Physics, University of Stavanger, NO-4036 Stavanger, Norway}
\affiliation{ Max-Planck-Institut f{\"u}r Gravitationsphysik (Albert-Einstein-Institut), Callinstra{\ss}e 38, 30167 Hannover, Germany}
\affiliation{Leibniz Universit{\"a}t Hannover, 30167 Hannover, Germany}

\author{Collin D. Capano}
\affiliation{Department of Physics, University of Massachusetts, Dartmouth, MA 02747, USA}
\affiliation{ Max-Planck-Institut f{\"u}r Gravitationsphysik (Albert-Einstein-Institut), Callinstra{\ss}e 38, 30167 Hannover, Germany}
\affiliation{Leibniz Universit{\"a}t Hannover, 30167 Hannover, Germany}

\author{Shilpa Kastha}
\affiliation{Niels Bohr International Academy, Niels Bohr Institute, Blegdamsvej 17, 2100 Copenhagen, Denmark}
\affiliation{ Max-Planck-Institut f{\"u}r Gravitationsphysik (Albert-Einstein-Institut), Callinstra{\ss}e 38, 30167 Hannover, Germany}
\affiliation{Leibniz Universit{\"a}t Hannover, 30167 Hannover, Germany}

\author{Alexander H. Nitz}
\affiliation{Department of Physics, Syracuse University, Syracuse, NY 13244, USA}

\author{Yi-Fan Wang}
\affiliation{ Max-Planck-Institut f{\"u}r Gravitationsphysik (Albert-Einstein-Institut), Am M{\"u}hlenberg 1, 14476 Potsdam, Germany}
\affiliation{ Max-Planck-Institut f{\"u}r Gravitationsphysik (Albert-Einstein-Institut), Callinstra{\ss}e 38, 30167 Hannover, Germany}
\affiliation{Leibniz Universit{\"a}t Hannover, 30167 Hannover, Germany}

\author{Julian Westerweck}
\affiliation{ Max-Planck-Institut f{\"u}r Gravitationsphysik (Albert-Einstein-Institut), Callinstra{\ss}e 38, 30167 Hannover, Germany}
\affiliation{Leibniz Universit{\"a}t Hannover, 30167 Hannover, Germany}
\affiliation{Institute for Gravitational Wave Astronomy and School of Physics and Astronomy, University of Birmingham, Edgbaston, Birmingham B15 2TT, United Kingdom}

\author{Alex B. Nielsen}
\affiliation{Department of Mathematics and Physics, University of Stavanger, NO-4036 Stavanger, Norway}

\author{Badri Krishnan}
\affiliation{Institute for Mathematics, Astrophysics and Particle Physics, Radboud University, Heyendaalseweg 135, 6525 AJ Nijmegen, The Netherlands}
\affiliation{ Max-Planck-Institut f{\"u}r Gravitationsphysik (Albert-Einstein-Institut), Callinstra{\ss}e 38, 30167 Hannover, Germany}
\affiliation{Leibniz Universit{\"a}t Hannover, 30167 Hannover, Germany}

\begin{abstract}

We study Bayesian inference of black hole ringdown modes for simulated binary black hole signals. We consider to what extent different fundamental ringdown modes can be identified in the context of black hole spectroscopy. Our simulated signals are inspired by the high mass event GW190521. We find strong correlation between mass ratio and Bayes factors of the subdominant ringdown modes.
The Bayes factor values and time dependency, and the peak time of the (3,3,0) mode align with those found analyzing the real event GW190521, particularly for high-mass ratio systems.

\end{abstract}

\maketitle

\section{Introduction\label{Introduction}}
The ringdown signal of a binary black hole (BBH) merger, which is related to the resonating space-time of the resulting final black hole \cite{Vishveshwara:1970cc, Vishveshwara:1970zz, Chandrasekhar:1975zza}, conveys extensive information about the post-merger object. Moreover, as we shall discuss in this paper, a detailed analysis of the ringdown signal allows one to infer properties of the pre-merger binary system.
At sufficiently late stages, the ringdown waveform can be simply represented as a superposition of quasinormal modes (QNMs).
The no-hair theorem in general relativity implies that the frequencies and damping times of the QNMs exclusively depend on the mass and spin of the single remnant black hole (BH). The amplitudes and phases of the QNMs depend on the initial pre-merger binary configuration. Systems with asymmetric progenitor masses tend to exhibit higher amplitudes of subdominant QNMs, while symmetric systems tend to have minimal amplitudes \cite{Wade:2013hoa,JimenezForteza:2020cve,Borhanian:2019kxt}.
The detection of multiple QNMs is often referred to as black hole spectroscopy.  
When multiple QNMs are detected simultaneously, they can be used not only to validate the no-hair theorem, but as we shall show in this paper, also to extract additional information about the mass ratio of the BBH.

In black hole spectroscopy, the detection of multiple ringdown modes enables us to assess the compatibility of a merger event with predictions from General Relativity (GR) \cite{Dreyer:2003bv, Kamaretsos:2011um}. 
The detection of multiple ringdown modes obeying these predictions serves as a highly compelling test of the underlying Kerr nature of the final black hole spacetime.

The QNMs are a representation of the post-merger waveform decomposed into a \emph{spheroidal} harmonic basis. The modes are enumerated by three integers $(\ell, m, n)$, satisfying the conditions $\ell\geq 2$, $-\ell\leq m \leq \ell$, and $n\geq 0$. Among these modes, those with $n\geq 1$ are commonly referred to as ``overtones.'' The integers $\ell$ and $m$ represent angular and azimuthal numbers, respectively, while $n$ indicates the overtone index.

Waveforms that model the entire signal through the inspiral, merger, and ringdown (IMR) phases of a binary instead use a \emph{spherical} harmonic basis to represent the signal. IMR waveforms have no overtones, and so their modes are characterized by just an $\ell$ and $m$. Consequently, whenever we use the notation $(\ell, m, n)$ in this paper --- e.g., (2,2,0) --- it corresponds to QNMs. When we use the notation $\ell m$ --- e.g., $22$ --- it signifies a mode of an IMR waveform model.

There are still several unresolved theoretical questions concerning black hole QNMs, which also have implications for observational studies. One of these questions relates to the starting time of the ringdown phase. After the formation of the remnant black hole, it initially possesses significant distortions from a Kerr black hole. As time progresses, these distortions dissipate; eventually, the black hole can be regarded as a linear perturbation of a Kerr black hole. The time at which this perturbative description becomes reliable, if at all, remains unclear \cite{Kamaretsos:2011um, Bhagwat:2017tkm,Thrane:2017lqn}. Additionally, studies exploring potential non-linear effects in the ringdown regime can be found in \cite{London:2014cma,Okounkova:2020vwu, Mitman:2022qdl, Lagos:2022otp, Cheung:2022rbm}.

The distinct regimes observed in a gravitational wave signal are anticipated to have corresponding behaviors in the strong-field dynamic spacetime region near the binary system \cite{Jaramillo:2012rr, Jaramillo:2011re, Prasad:2020xgr}. Investigations into black hole horizon geometry during the post-merger phase have shown the potential identification of a ringdown regime using horizon dynamics \cite{Mourier:2020mwa, Forteza:2021wfq, Pook-Kolb:2020jlr, Gupta:2018znn, Chen:2022dxt}. Recent work also shows the presence of quadratic non-linearities at the horizon \cite{Khera:2023lnc}. 
The determination of the final black hole parameters based solely on the ringdown signal is sensitive to the assumed starting time of the ringdown \cite{Kamaretsos:2011um,Carullo:2018sfu,Carullo:2019flw,Cotesta:2022pci}. Different choices of the starting time can lead to distinct outcomes \cite{Cabero:2017avf, Kastha:2021chr}. An analysis focused solely on the ringdown phase needs to explicitly exclude earlier portions of the signal where the perturbative description is not valid.

It has been widely anticipated that the current generation of gravitational wave detectors would predominantly detect the most prominent ringdown mode \cite{Berti:2016lat, Cabero:2019zyt}. This expectation was rooted in astrophysical assumptions concerning the mass distributions and mass ratios of binary black hole systems across the observable universe, which directly impact the amplitudes of different ringdown modes \cite{Wade:2013hoa,Borhanian:2019kxt,JimenezForteza:2020cve}. 
Nevertheless, evidence supporting the existence of an overtone accompanying the dominant mode of GW150914 was provided in \cite{Isi:2019aib, Giesler:2019uxc,Crisostomi:2023tle}. These studies demonstrated the feasibility of modeling the gravitational waveform as a combination of ringdown modes, even starting in the merger phase, by incorporating overtones of the dominant mode. However, several questions related to both the data analysis methods used to detect overtones and the validity of the model at merger remain unresolved~\cite{Cotesta:2022pci, Isi:2022mhy,Baibhav:2023clw}. The stability of overtones under small perturbations has also been widely studied \cite{Nollert:1996rf, Nollert:1998ys, Jaramillo:2020tuu, Jaramillo:2021tmt, Destounis:2021lum}.

GW190521 was detected by the Advanced LIGO and Advanced Virgo detectors \cite{LIGOScientific:2020iuh}. Although a multitude of scenarios have been put forward regarding the source of this event \cite{Bustillo:2020syj,Romero-Shaw:2020thy,Gayathri:2020coq,Abedi:2021tti, Wang:2021gqm,Gamba:2021gap, DallAmico:2021umv,Shibata:2021sau,Abedi:2022bph,CalderonBustillo:2022cja}, the most conservative explanation is that it was caused by the merger of a quasi-circular BBH~\cite{LIGOScientific:2020iuh, LIGOScientific:2020ufj}.
Assuming GW190521 is a BBH merger, its inferred total mass would categorize it as one of the most massive binary black hole systems observed thus far \cite{LIGOScientific:2021djp, Nitz:2021zwj}; other interpretations have even proposed higher total masses \cite{Bustillo:2020syj}. The substantial mass implies that a significant portion of the inspiral phase occurs below the sensitive frequency range of the detectors, meaning the recorded signal is dominated by the post-inspiral merger and ringdown phases. Consequently, focusing the analysis on the ringdown phase becomes particularly valuable as it circumvents certain challenges associated with modeling the inspiral phase of the progenitor system.

While the nature of the progenitors of the GW190521 event remains open, the most likely outcome in most scenarios is the formation of a single black hole. The event GW190521 exhibits clear evidence of a prominent ringdown mode emitted by the final black hole following the merger \cite{LIGOScientific:2020iuh}. In the study by Capano et al. \cite{Capano:2021etf}, an additional sub-dominant ringdown mode was identified in the signal. The dominant mode is consistent with the $(\ell,m,n)=(2,2,0)$ ringdown mode of a Kerr black hole, while the second mode is consistent with the sub-dominant fundamental $(\ell,m,n)=(3,3,0)$ mode. As detailed in \cite{Capano:2021etf}, under the assumption of a Kerr black hole, the Bayes factor favoring the existence of both the $(2,2,0)$ and $(3,3,0)$ modes over just the $(2,2,0)$ mode or the combination of $(2,2,0)$ and $(2,2,1)$ modes is estimated to be $\eventbf{}$.
Subsequent work \cite{Capano:2022zqm} based on simulated signals and real detector noise showed that the probability of falsely detecting these two modes when only one is present in a quasi-circular binary is $\sim 0.02$. An additional agnostic analysis assumed no specific relationship between the ringdown modes. This indicated that only 1 in 250 simulated signals without a (3,3,0) mode produces a result at least as significant as the one observed for GW190521 \cite{Capano:2022zqm}.

A recent ringdown study by Siegel et al. (2023)~\cite{Siegel:2023lxl} finds hints of multiple observable QNMs in GW190521. 
In their analysis, they find evidence for the presence of the $(2,1,0)$ mode with an amplitude comparable to, and even exceeding, the $(2,2,0)$ mode. They also find evidence for a higher-frequency subdominant mode, in agreement with Capano et al.~\cite{Capano:2021etf}. However, they identify this mode as the $(3,2,0)$ QNM instead of the $(3,3,0)$ mode, as in Capano et al. Their result aligns more closely with results obtained from the numerical relativity surrogate model NRSur7dq4~\cite{Varma:2019csw}. In particular, they ascribe the large amplitude of the $(2,1,0)$ to large precession in the binary.

\begin{figure*}[hbt!]
    \centering
    \begin{minipage}[b]{0.49\textwidth}
        \centering
        \includegraphics[width=\textwidth]{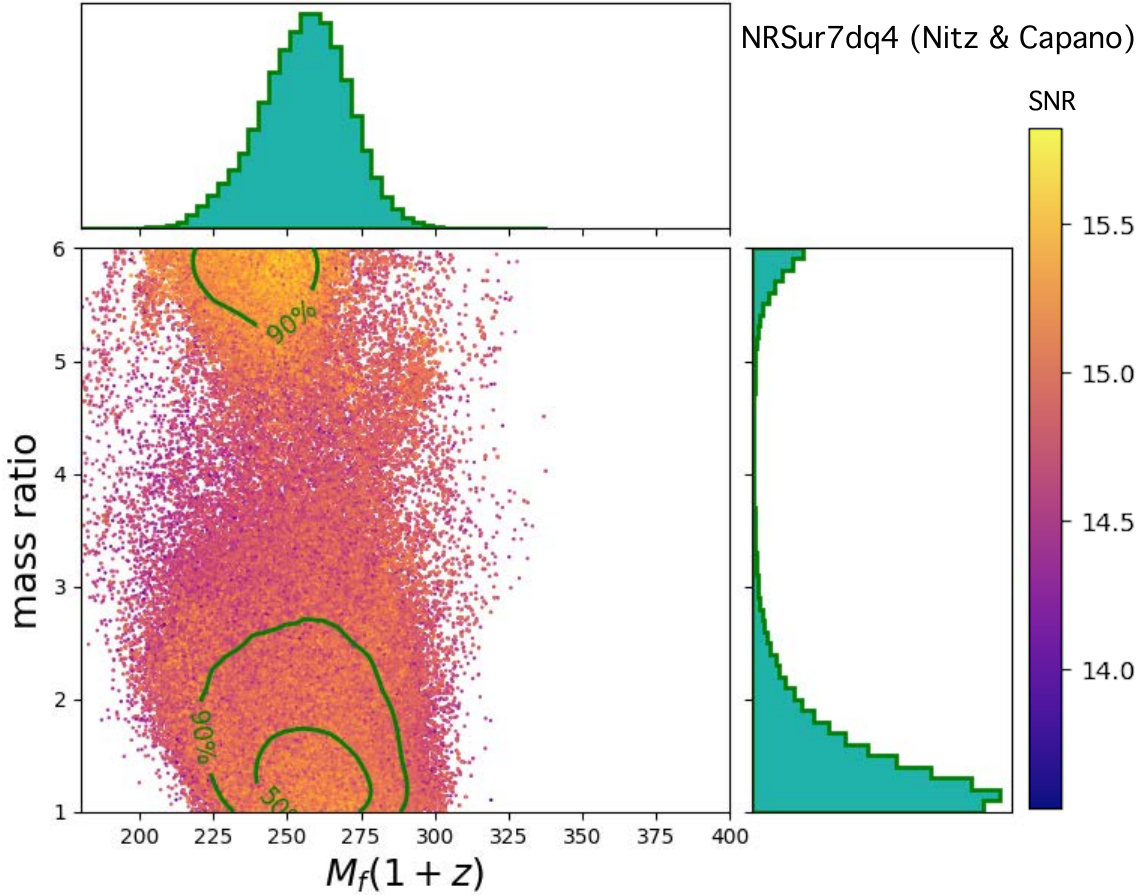}
    \end{minipage}
    \begin{minipage}[b]{0.49\textwidth}
        \centering
        \includegraphics[width=\textwidth]{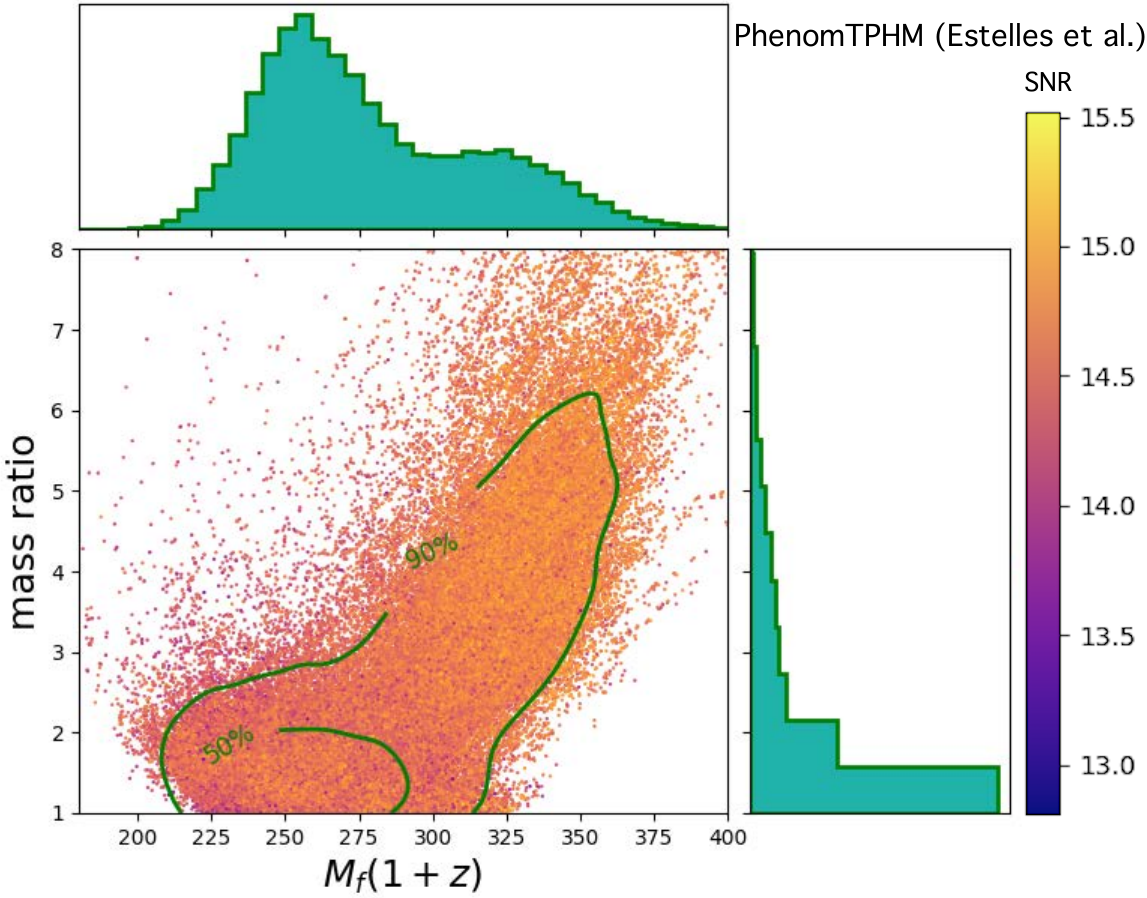}
    \end{minipage}
    \caption{A comparison of the final mass, mass ratio and SNR of samples from IMR analyses of GW190521, using the NRSur7dq4~\cite{Nitz:2020mga} (\textit{Left}) and IMRPhenomTPHM (\textit{Right})\cite{Estelles:2021jnz} waveform model. The solid lines in the plots represent the $50\%$ and $90\%$ credible contours. The dots' color represents the samples' SNR. Notably, the NRSur7dq4 results exhibit a second mode with higher SNR at larger mass ratio, which agrees with the findings from the ringdown analysis.}
    \label{fig:imrcompare-time_mass_ratio_final_mass}
\end{figure*}

Even under the assumption that GW190521 was caused by the merger of a quasi-circular BBH, estimates of its mass ratio vary widely depending on the waveform model used and the prior assumed. In \cite{LIGOScientific:2020iuh}, the binary was reported as being approximately equal mass. That analysis used the IMR model NRSur7dq4 (which is a surrogate model derived from numerical relativity simulations~\cite{Varma:2019csw}) and a prior uniform in component masses. However, a subsequent analysis by Nitz \& Capano~\cite{Nitz:2020mga} using a prior uniform in mass ratio $m_1/m_2$ (where $m_1 \geq m_2$) found support for larger mass ratios. Using this prior, Nitz \& Capano found a bimodal posterior distribution in the mass ratio when using NRSur7dq4, with the second mode railing against $m_1/m_2 = 6$, which is the upper bound of the valid range for NRSur7dq4~\cite{Varma:2019csw}. The posterior from this analysis is replicated in the left panel of Fig.~\ref{fig:imrcompare-time_mass_ratio_final_mass}. Notably, the signal-to-noise ratio of the mass ratio $\sim 6$ points are slightly larger than the equal-mass points. This indicates that the equal-mass scenario favored in \cite{LIGOScientific:2020iuh} is largely due to the choice of prior.

A further analysis of GW190521 by Estelles et al.~\cite{Estelles:2021jnz} using the IMR model PhenomTPHM also found support for larger mass ratios, depending on the mass prior chosen. That result, which is reproduced in the right panel of Fig.~\ref{fig:imrcompare-time_mass_ratio_final_mass}, where a uniform prior in component masses
(in detector frame) with a range $m_{1,2}^{det} \in [10, 400]M_{\odot}$ and default mass ratio $m2/m1$ constraints $\in [0.035, 1.0]$ is employed, also yielded a bimodal posterior, with one mode at equal mass and another as mass ratio $\sim5$. The maximum SNR was about equal between the two. The larger mass ratio mode from the PhenomTPHM results yielded a final mass and spin estimate consistent with that found in Capano et al., whereas the NRSur7dq4 results (even at mass ratio $\sim6$) did not~\cite{Capano:2022zqm}.

The discrepancy in mass ratio estimates is due to the challenging nature of GW190521. Only about one cycle is observable above noise prior to merger. In order to match both that cycle and the ringdown, the IMR models need to go to high precession and/or larger mass ratio. However, large precession, mass ratios larger than $\sim2$, and the transition from late inspiral to ringdown is one of the hardest areas for waveform models to capture, as there are relatively few NR simulations in this space to calibrate against. For example, the NRSur7dq4 approximant is calibrated to NR simulations with mass ratios up to $4$ and dimensionless spin magnitudes $<0.8$, and was validated for mass ratios up to $6$ by performing mismatch calculations to simulations with total masses up to $200\,\mathrm{M}_\odot$~\cite{Varma:2019csw}. For comparison, GW190521 is best matched by the NRSur7dq4 model with in-plane spin on the large object of $\gtrsim0.8$, mass ratio $\sim 6$, and total mass $>250\,\mathrm{M}_\odot$. Conversely, PhenomTPHM is calibrated to aligned-spin simulations, relying on a ``twisting-up'' procedure that may be less reliable late in the inspiral \cite{Estelles:2021jnz}.

To better understand the discrepancy in mass ratio estimates for GW190521, in this paper we use QNM templates to analyze a range of different simulated signals. We invert the typical question of parameter estimation: instead of asking which IMR template best matches (the largely ringdown dominated) GW190521, we ask which QNM models best match the post-merger of an IMR simulation. To that end, 
we analyze simulated black hole ringdown signals containing both the dominant $(2,2,0)$ mode and various combinations of sub-dominant modes, including both fundamental modes and first overtones. The full set of mode combinations considered is listed in Table(\ref{Table 1}).

\begin{table}
\label{Table 1}
\begin{tabular}{|l|}
  \hline
  Combinations of Ringdown Modes \\
  \hline
  (2,2,0) \\
  (2,2,0)+(2,2,1) \\
  (2,2,0)+(2,2,1)+(3,3,0) \\
  (2,2,0)+(2,2,1)+(3,3,0)+(2,1,0) \\
  (2,2,0)+(2,2,1)+(3,3,0)+(3,2,0) \\
  (2,2,0)+(2,2,1)+(3,3,0)+(2,1,0)+(3,2,0) \\
  (2,2,0)+(2,2,1)+(2,1,0) \\
  (2,2,0)+(3,3,0) \\
  (2,2,0)+(3,3,0)+(2,1,0) \\
  (2,2,0)+(3,3,0)+(2,1,0)+(3,2,0) \\
  (2,2,0)+(3,3,0)+(3,2,0) \\
  \hline
\end{tabular}
 \caption{Combinations of black hole ringdown modes analyzed in this work.}
 \end{table}

Our simulated signals use the IMR waveform model IMRPhenomTPHM \cite{Estelles:2021gvs}. In order to better understand the underlying physics of GW190521, this study employs zero-noise injections to mitigate complexities introduced by real detector noise. The data we analyze therefore consists of only the signal as it would be seen in the detectors without noise being present. This in effect samples over a large population of noise realisations from the assumed Gaussian distribution. While accounting for the detector response, we can then focus solely on the intrinsic characteristics of the signal.


The IMRPhenomTPHM injections are drawn from posterior samples provided in Ref.~\cite{Estelles:2021jnz}. We investigate the compatibility of BBH waveforms with the presence of subdominant QNM modes. Our findings indicate that within this family of quasi-circular binary systems, previously obtained QNM results favor an asymmetric system for the GW190521 event.


\begin{figure}
    \includegraphics[width=\columnwidth]{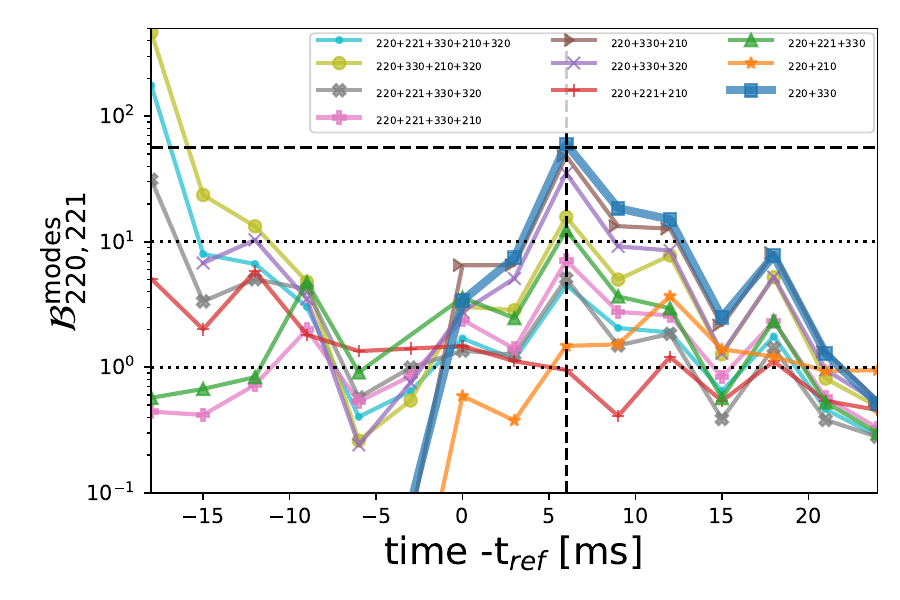}
    \caption{Bayes factors $\mathcal{B}^{\text{modes}}_{220,221}$ for various models with the indicated modes in real GW190521 data. These are calculated relative to the stronger of the two models containing either only the $(2,2,0)$ or both the $(2,2,0) + (2,2,1)$ modes. The dashed lines show where the (3,3,0) mode peaks. As depicted in this figure, the Bayes factor strongly supports the presence of both the $(2,2,0)$ and $(3,3,0)$ modes over either the $(2,2,0)$ mode alone or a combination of the $(2,2,0)$ and $(2,2,1)$ modes, with an estimated value of $\eventbf{}$ at $\eventbftime$.}
    \label{fig:gw190521}
\end{figure}

\begin{figure*}
    \centering
    \includegraphics[width=0.8\textwidth]{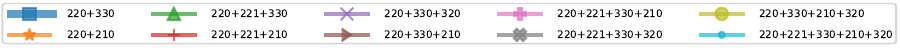}
    \begin{minipage}[b]{0.42\textwidth}
        \centering
        \includegraphics[width=\textwidth]{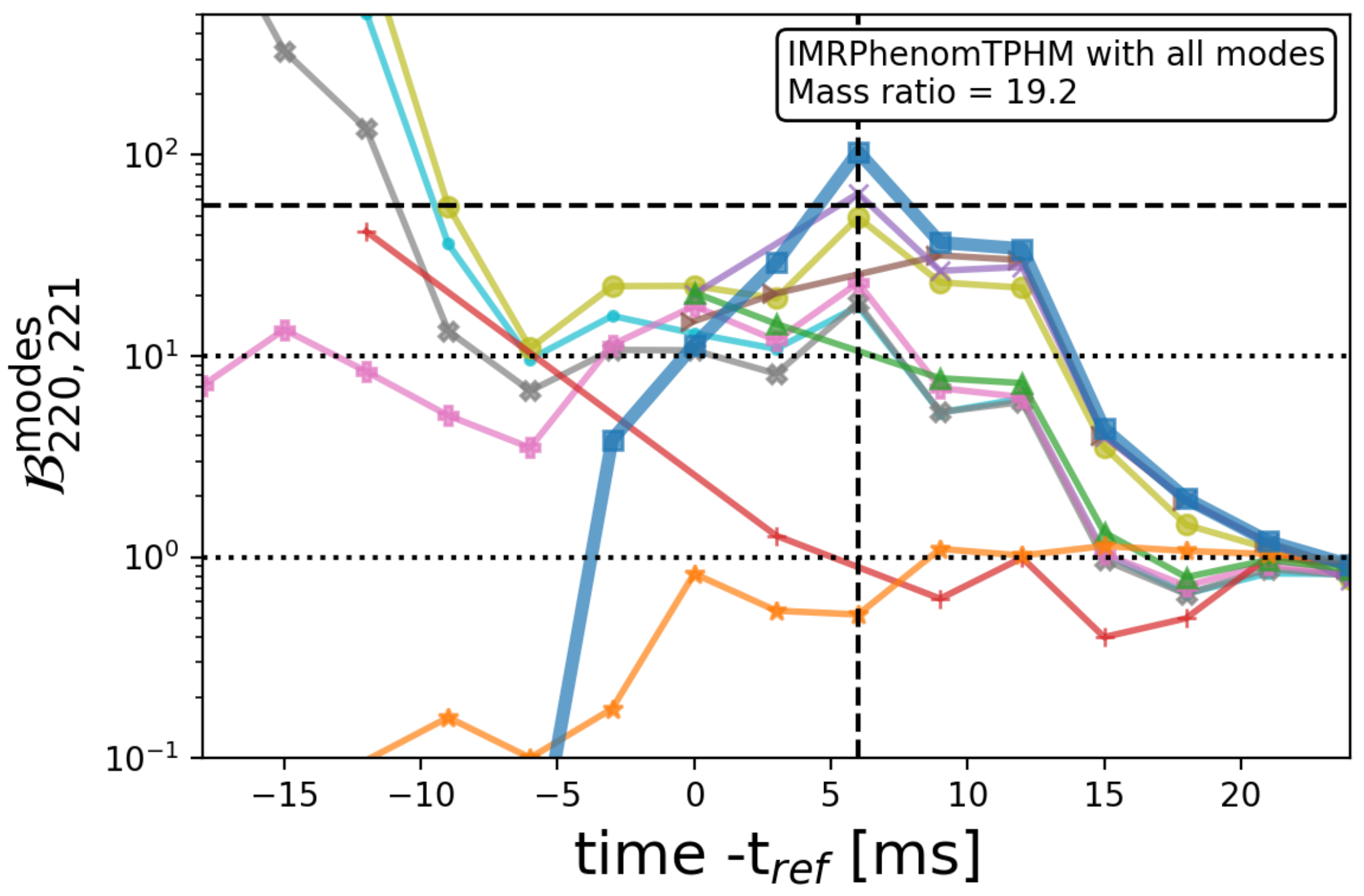}
    \end{minipage}
    \begin{minipage}[b]{0.42\textwidth}
        \centering
        \includegraphics[width=\textwidth]{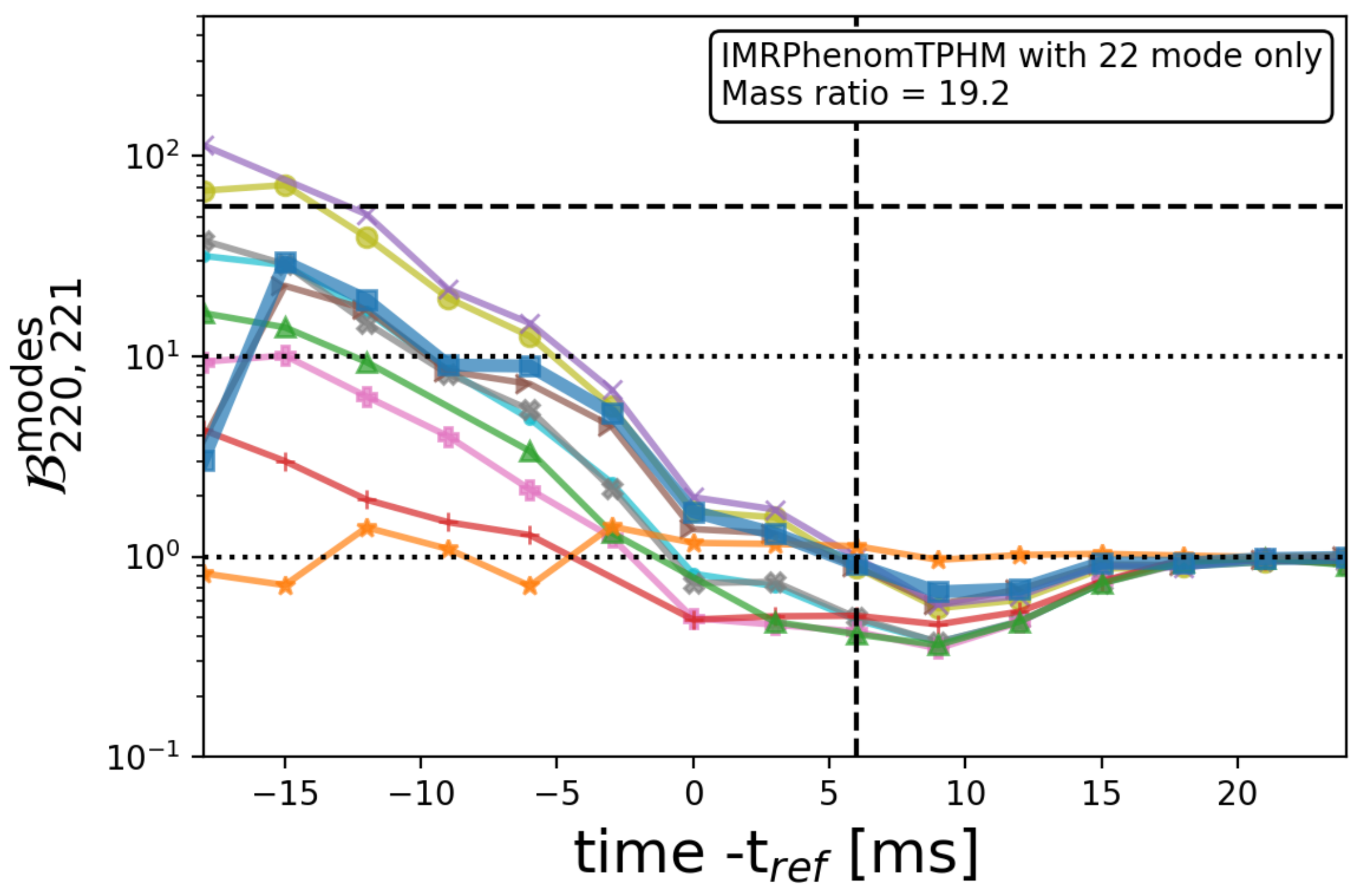}
    \end{minipage}
    
    \begin{minipage}[b]{0.42\textwidth}
        \centering
        \includegraphics[width=\textwidth]{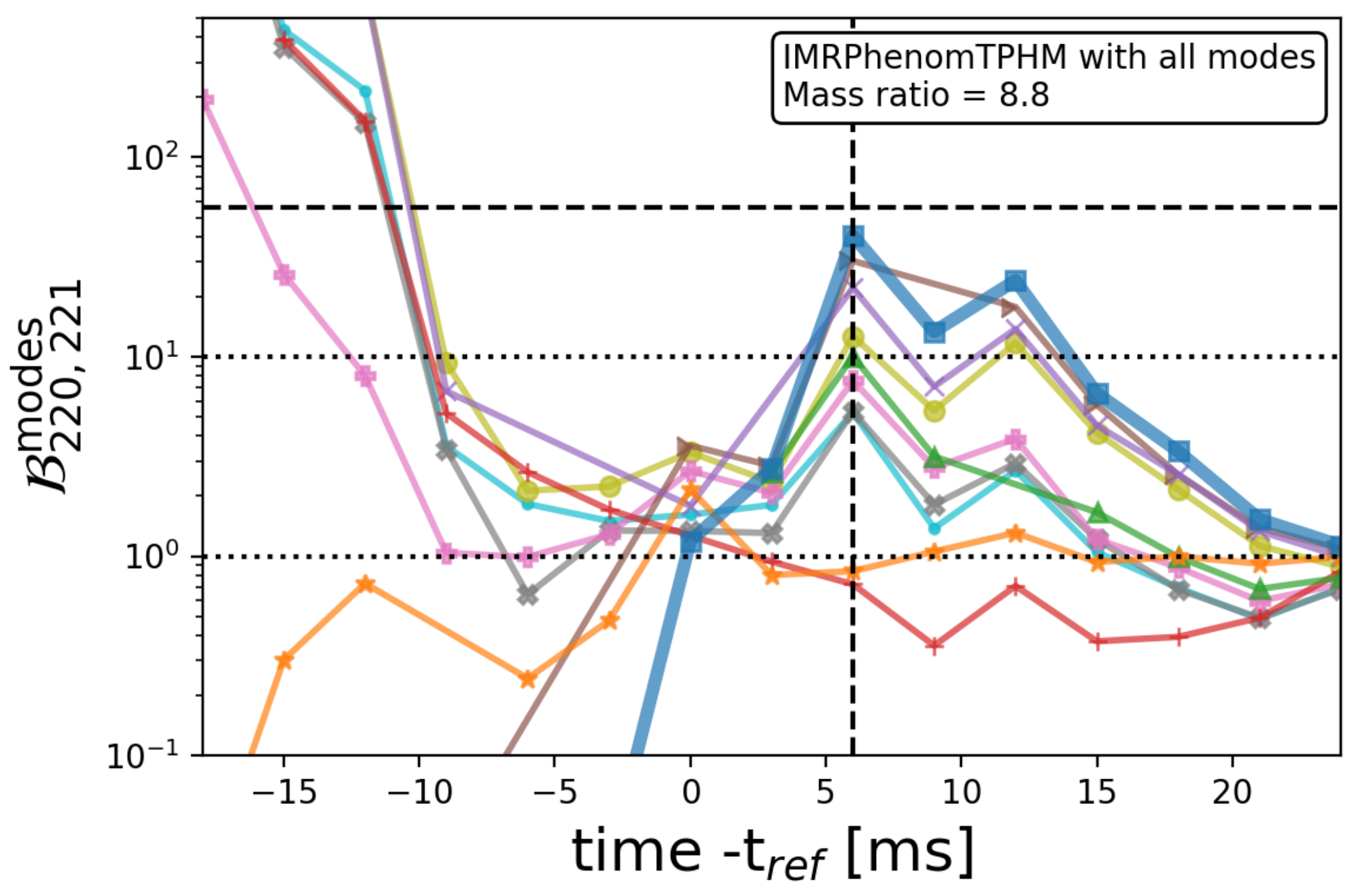}
    \end{minipage}
    \begin{minipage}[b]{0.42\textwidth}
        \centering
        \includegraphics[width=\textwidth]{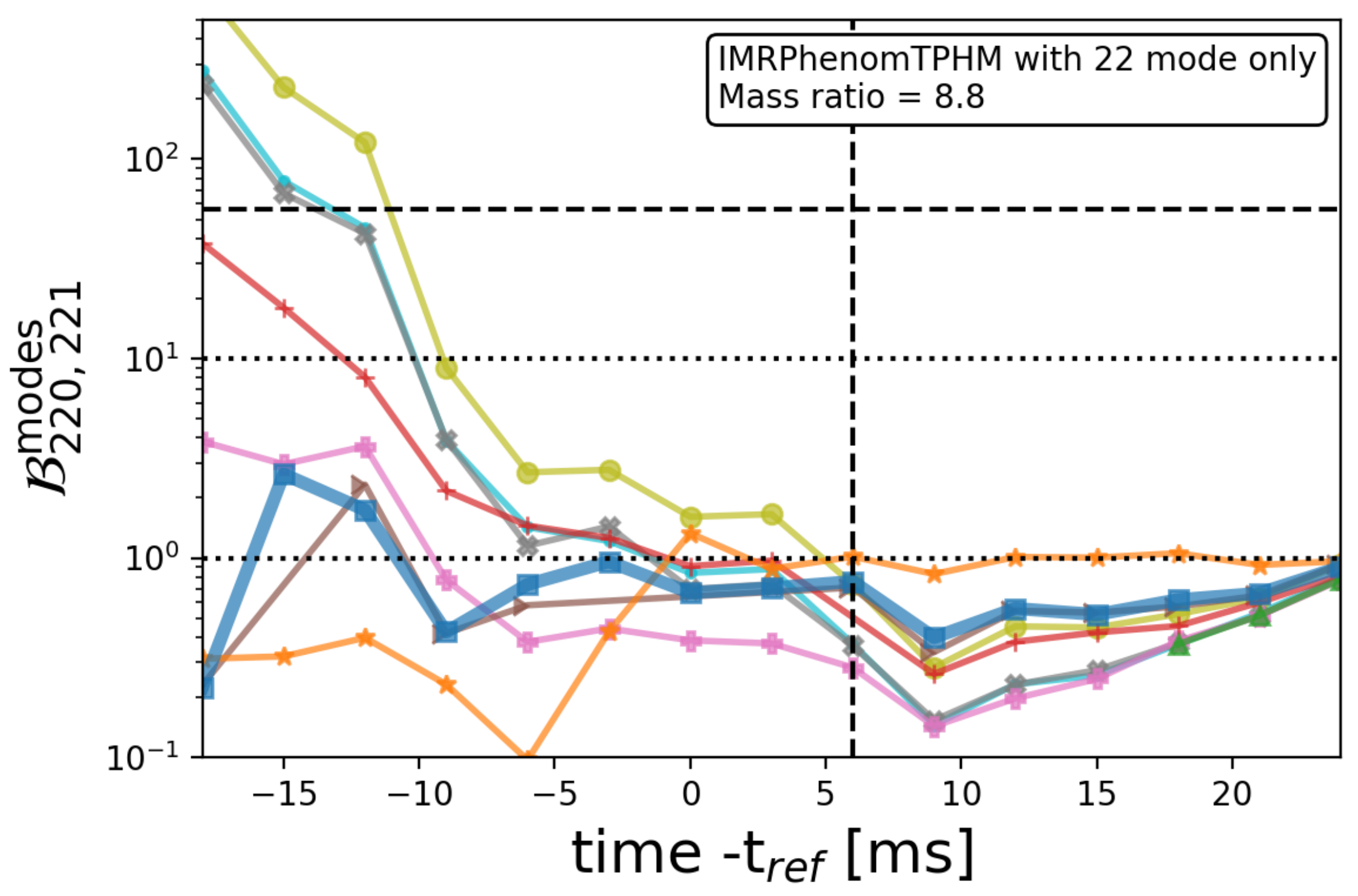}
    \end{minipage}

    \begin{minipage}[b]{0.42\textwidth}
        \centering
        \includegraphics[width=\textwidth]{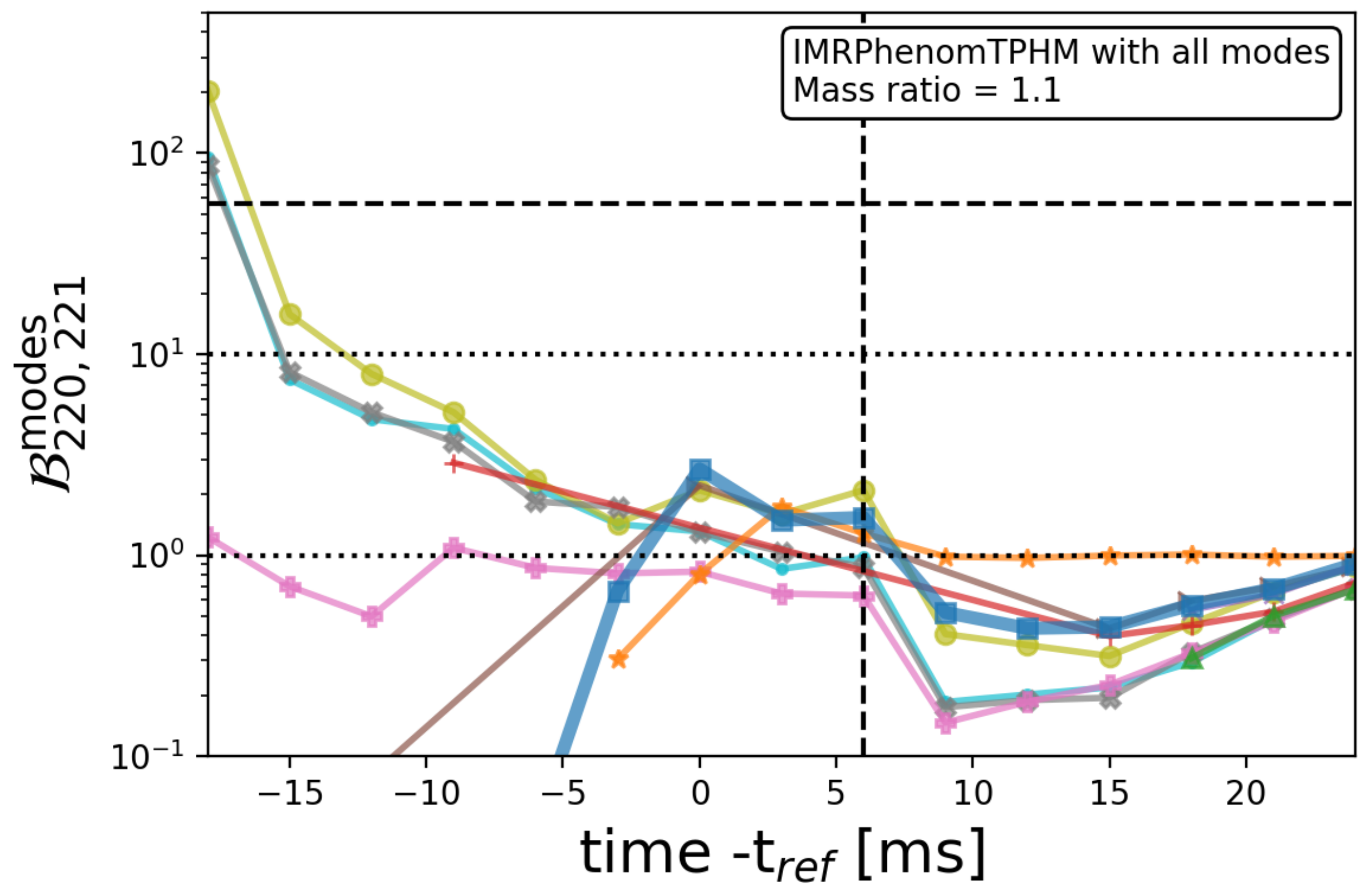}
    \end{minipage}
    \begin{minipage}[b]{0.42\textwidth}
        \centering
        \includegraphics[width=\textwidth]{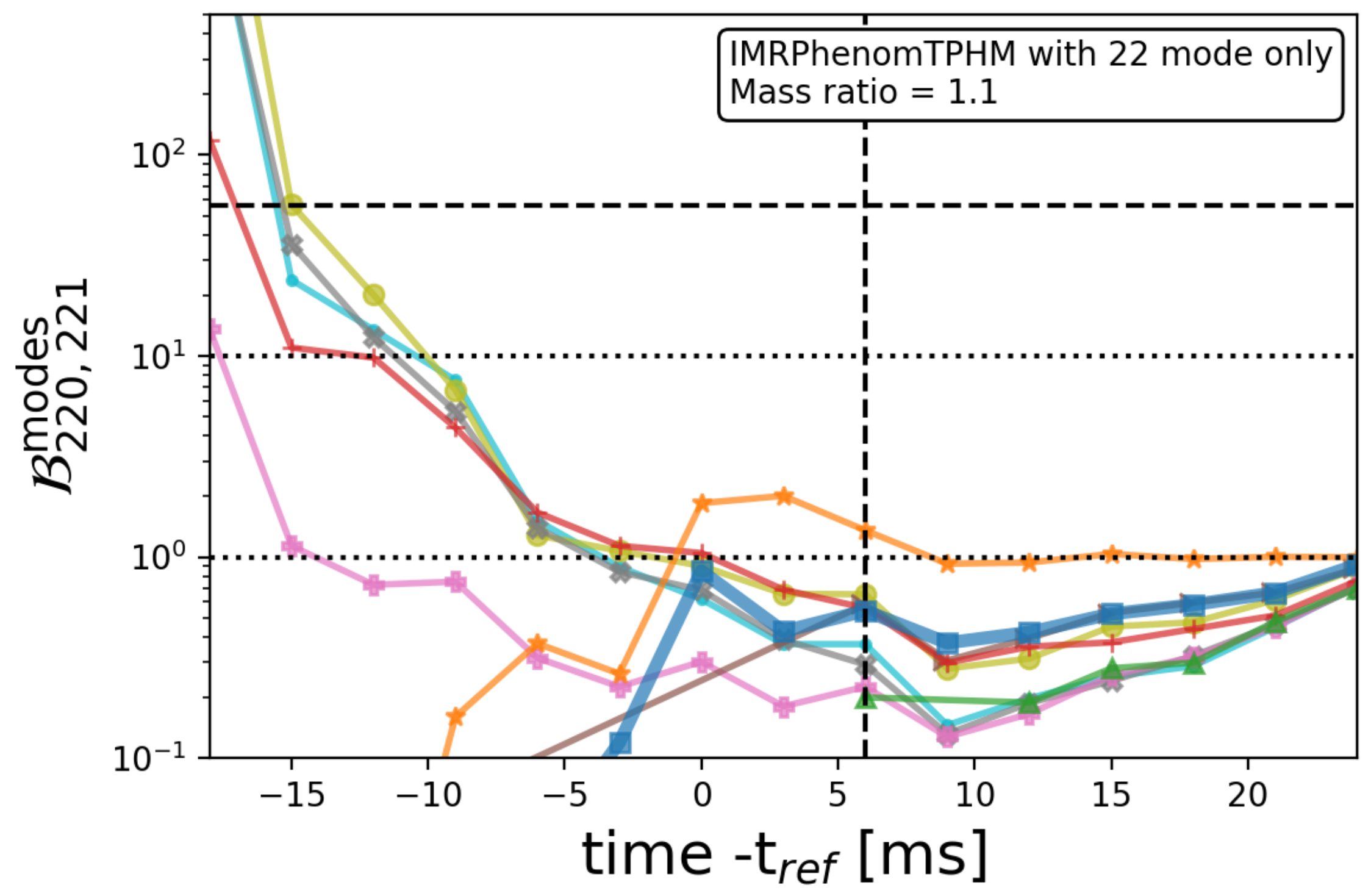}
    \end{minipage}
    \caption{
    Results of the ringdown analysis for simulated signals of set \emph{A} in terms of Bayes factors as shown in Fig.~\ref{fig:gw190521}.
    Set \emph{A} consists of 100 injections drawn from the GW190521 IMR-posterior samples. The injected signals use the IMRPhenomTPHM waveform model. 
    Both columns show the same injection parameters, but either using all modes implemented in IMRPhenomTPHM (left, \emph{\signalinj{}} subset) or only the 22 mode (right, \emph{\controlinj{}} subset).
    Top row: Highest mass ratio injection, $q=19.2$ Center row: Injection with $(3,3,0)$ Bayes factor closest to that for GW190521, $q=8.8$. Bottom row: Lowest mass ratio injection of set \emph{A}, $q=1.1$. The ringdowm analysis method and selection of modes are identical for all results shown. The dashed lines show where the (3,3,0) mode Bayes factor for GW190521 peaks in Fig.~\ref{fig:gw190521}. This figure illustrates that as the mass ratio increases within the \emph{\signalinj} injection set (left panel), our analysis patterns of multiple modes—encompassing peak time, Bayes factor values, etc.—increasingly mirrors the pattern seen in GW190521 (illustrated in Fig.~\ref{fig:gw190521}). Conversely, when we conduct an equivalent analysis on the \emph{\controlinj} injections (right panel), featuring only the 22 mode, we find no similarity to the observed pattern.}
    \label{fig:QNMsfor100injections}
\end{figure*}

\begin{figure*}
    \centering
    \begin{minipage}[b]{0.49\textwidth}
        \centering
        \includegraphics[width=\textwidth]{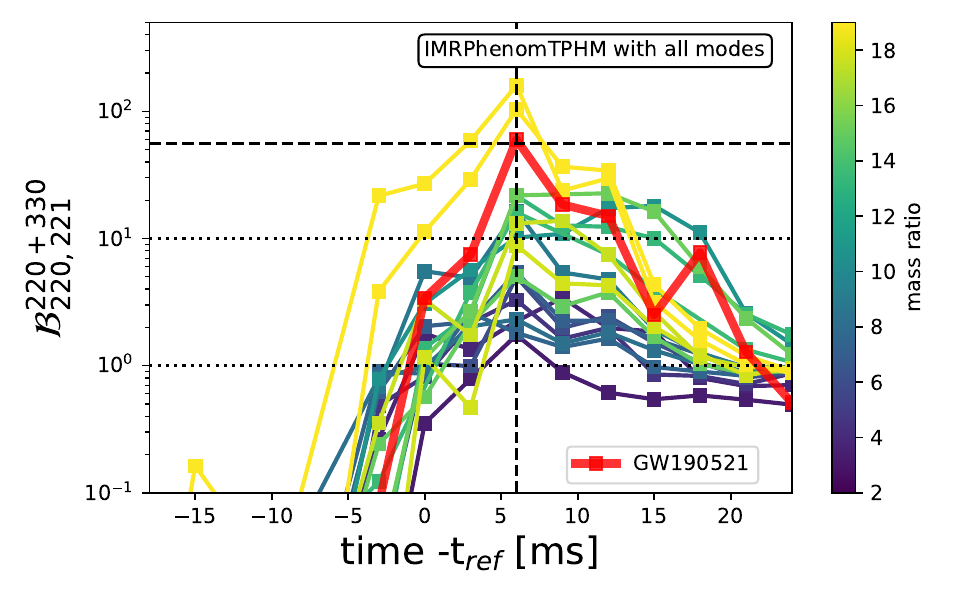}
    \end{minipage}
    \begin{minipage}[b]{0.49\textwidth}
        \centering
        \includegraphics[width=\textwidth]{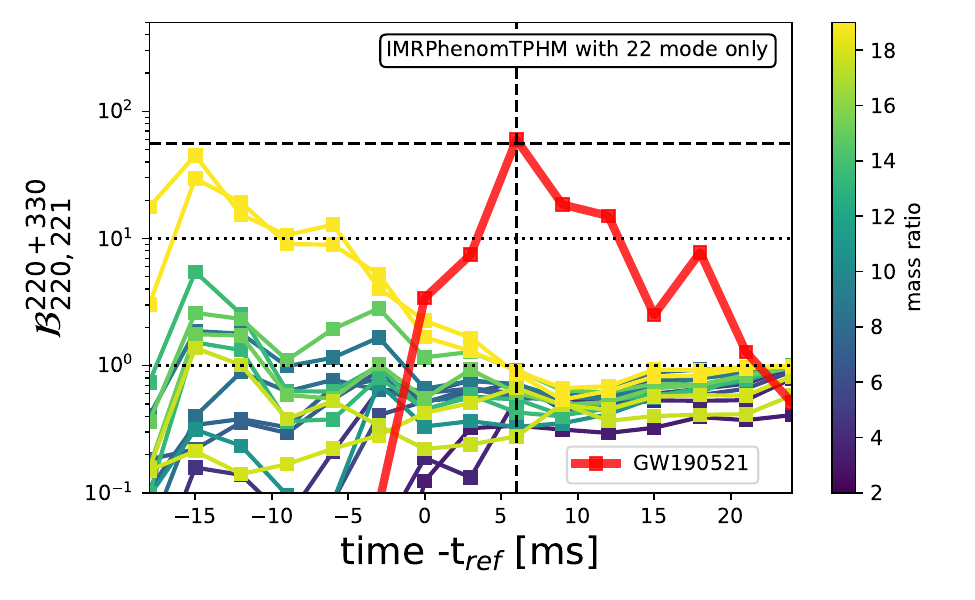}
    \end{minipage}
    \caption{
    Results of the ringdown analysis of set \emph{A}, using as templates the $(2,2,0) + (3,3,0)$ quasi-normal modes. Set \emph{A} has 100 injections drawn from the GW190521 IMR-posterior samples which use the IMRPhenomTPHM waveform model. The injections' mass ratio ranges from $q=1.1$ to $q=19.2$. For visibility, we display only 17 randomly selected injections of the total 100, uniformly distributed in mass ratio.
    Injections for the left panel have all modes of IMRPhenomTPHM, while those for the right panel have only the 22 mode.
    The ringdowm analysis method and template modes are identical for left and right. The red curve represents the result from the $(2,2,0)+(3,3,0)$ analysis for GW190521 and dashed lines show where its Bayes factors peak. In the left panel, this figure demonstrates that as the mass ratio increases in the \emph{\signalinj} injections, our analysis pattern—encompassing the peak time, Bayes factor values, etc.—for $(2,2,0)+(3,3,0)$ progressively aligns with the pattern observed in GW190521 (illustrated in red). In contrast, conducting a similar analysis on the \emph{\controlinj} injections (right panel), which feature only the 22 mode, reveals no resemblance to the observed pattern. This figure prominently demonstrates the robust support from the Bayes factor for the existence of both $(2,2,0)+(3,3,0)$ modes over either $(2,2,0)$ or $(2,2,0)+(2,2,1)$ modes at $\sim\eventbftime$ specifically for high mass ratio systems.}
    \label{fig:bfvstime}
\end{figure*}

\begin{figure*}
    \centering
    \begin{minipage}[b]{0.49\textwidth}
        \centering
        \includegraphics[width=\textwidth]{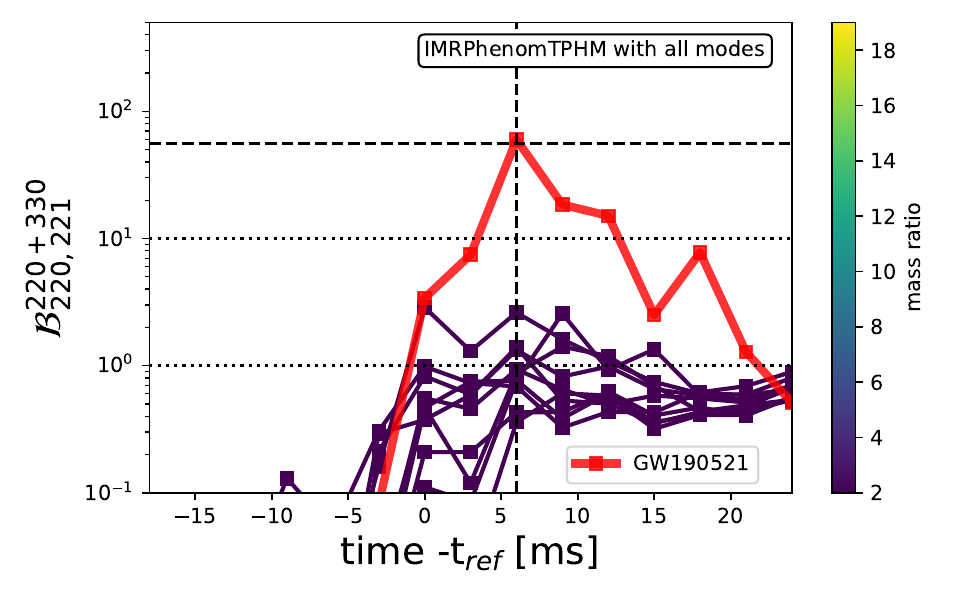}
    \end{minipage}
    \begin{minipage}[b]{0.49\textwidth}
        \centering
        \includegraphics[width=\textwidth]{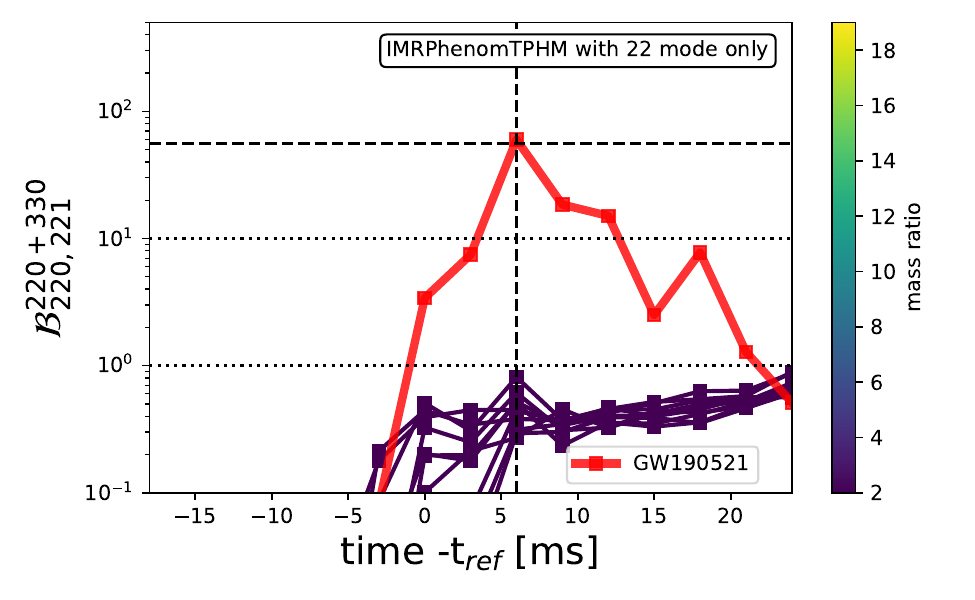}
    \end{minipage}
    \caption{The same as Fig.~\ref{fig:bfvstime} for the 10 injections of set \emph{B}. These are low mass-ratio injections (set \emph{B}), and neither the \emph{\signalinj} (left panel) nor the \emph{\controlinj} (right panel) injections exhibit similarity to the pattern observed in GW190521 (illustrated in red).
    For these low mass-ratio injections, the Bayes factors indicate no support for the existence of both $(2,2,0)+(3,3,0)$ modes.}
    \label{fig:bfvstimeqlt2}
\end{figure*}

\begin{figure*}
    \centering
    \begin{minipage}[b]{0.49\textwidth}
        \centering
        \includegraphics[width=\textwidth]{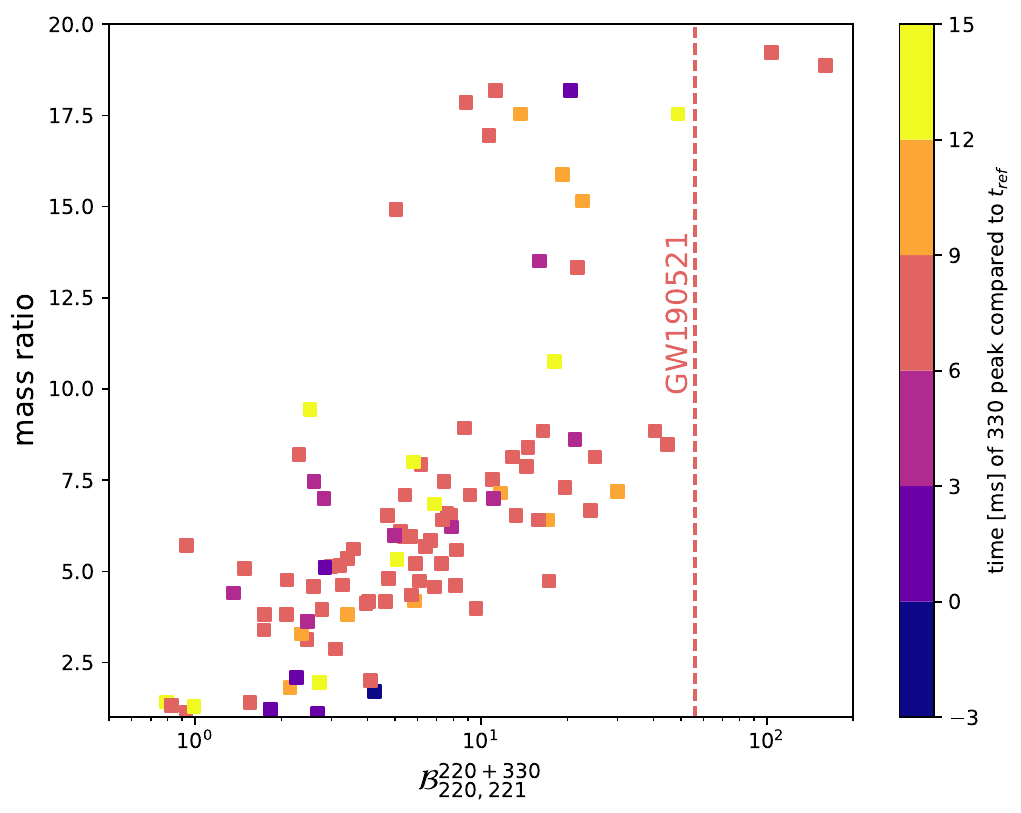}
    \end{minipage}
    \begin{minipage}[b]{0.49\textwidth}
        \centering
        \includegraphics[width=\textwidth]{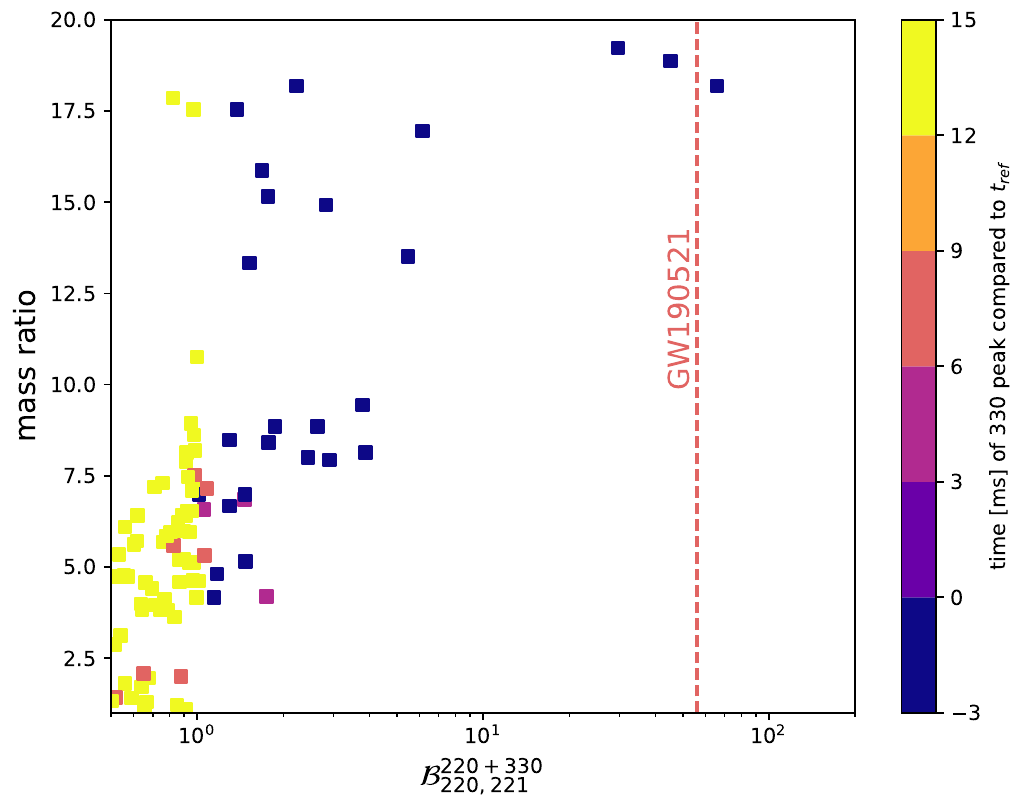}
    \end{minipage}
    \caption{
    Set \emph{A} simulated signals: Bayes factors $\mathcal{B}^{220+330}_{220,221}$ for the $(2,2,0)+(3,3,0)$ model, maximized over time, vs.\ mass ratio of the corresponding injection. Bayes factors are again calculated with respect to the stronger of the two models consisting of either the $(2,2,0)$ or the $(2,2,0) + (2,2,1)$ modes. The vertical dashed line shows the Bayes factor for GW190521. The colorbar represents the time of the peak $(3,3,0)$ Bayes factor relative to the reference time. \textit{Left}: Injections with all IMR modes (\emph{\signalinj}). \textit{Right}: The same injections with only the 22 IMR mode (\emph{\controlinj}). For high mass-ratio systems (set \emph{A}), the peak time and Bayes factor values for the \emph{\signalinj} injections (left panel) are shifted towards those obtained for GW190521 (dashed vertical line), compared to those of the \emph{\controlinj} injections (right panel).}
    \label{fig:bfvsq}
\end{figure*}



The key results for GW190521 in real data for various combinations of modes are shown in Fig.~\ref{fig:gw190521}. 
Figs.~\ref{fig:QNMsfor100injections} and \ref{fig:bfvstime} illustrate the outcome of a similar analysis as depicted in Fig.~\ref{fig:gw190521}, but applied to simulated signals (set \emph{A}\footnote{For definition of set \emph{A} and set \emph{B}, please refer to Section \ref{sec:injectionsets}.}) that exhibit a relatively high amplitude of the (3,3,0) mode (left panel) and simulated signals containing only the 22 IMR mode (right panel). Fig.~\ref{fig:bfvstimeqlt2} represents the same analysis as shown in Fig.~\ref{fig:bfvstime} while applied to simulated signals in set \emph{B}. The plots in Figs.~\ref{fig:QNMsfor100injections} are arranged in order based on different mass ratios (only for three injections), with each row representing a distinct mass ratio. Fig.~\ref{fig:bfvsq} represents the relation between the maximum Bayes factor for the (3,3,0) mode and the initial binary's mass ratio.

In this paper, we analyze data within a time span of 42 milliseconds that includes both the pre-merger and post-merger regimes. Therefore, through our investigation of the quasinormal modes (QNMs), we are able to identify the point at which QNMs best model the signal.


In this paper we utilized a range of starting times for the ringdown analysis that covers a range even larger than full range of possible uncertainty that may occur in our study. This range is even larger than what we investigated in our previous studies \cite{Capano:2021etf,Capano:2022zqm}. Further details can be found in Section \ref{sec:injectionsets}.

In Section \ref{sec:ringdownPE}, we provide supplementary information regarding the data treatment in this paper. The methodology employed in this study is similar to the one utilized in Capano et al. \cite{Capano:2021etf,Capano:2022zqm}, with the exception of an extended time span of data considered, the utilization of zero-noise data, and the inclusion of additional modes not considered in those works. Section \ref{sec:injectionsets} elucidates the procedure for generating the simulated data sets (\emph{A}, \emph{B}, \emph{\signalinj}, and \emph{\controlinj}) for an event similar to GW190521, as well as the definition of the modified Bayes factor.
Finally, in results and discussions \ref{sec:conclusions}, we present the results obtained in this study and provide a concise summary of these findings.


\section{Ringdown model and search pipeline}
\label{sec:ringdownPE}

In this section, we provide a summary of our data analysis procedure, focused exclusively on analyzing the ringdown component portion of the complete signal. 
Here, important challenges are the accurate identification of the ringdown section of the data, and restricting the analysis to this section while removing correlations with data outside the desired range.

The ringdown waveform model is described by the mathematical expression or functional form
\begin{equation}
\label{linear_spectrum}
h_+ + i h_\times = {\displaystyle \frac{M_f}{D_L}} \sum_{\lmn} {}_{_{-2}}S_{\lmn} (\iota, \varphi,{\chi}_f) 
A_{\lmn} e^{i(\Omega_{\lmn}t +  \phi_{\lmn})} \, .
\end{equation}
The waveform model for the ringdown can be expressed in terms of the plus ($h_+$) and cross ($h_\times$) polarizations of the gravitational wave. It depends on parameters such as the total mass of the remnant black hole in the detector frame $M_f$ and the source luminosity distance $D_L$. The waveform is decomposed using the spin-2 weighted spheroidal basis ${}_{-2}S_{\lmn}$, which is a function of the remnant black hole's spin $\chi_f$, the inclination angle $\iota$, and the azimuthal angle $\varphi$ relative to the line-of-sight to the observer. The amplitude and phase of each quasi-normal mode is represented by $A_{\lmn}$ and $\phi_{\lmn}$, respectively.
The complex frequency of the ringdown waveform is defined as $\Omega_{\lmn} = 2\pi f_{\lmn} + i/\tau_{\lmn}$. The characteristic frequency $f_{\lmn}$ and decay time $\tau_{\lmn}$ are solely determined by the mass and spin of the remnant black hole, as predicted by the no-hair theorem in general relativity.

Our analysis must be restricted to the data where the QNM waveform model is valid. 
This is achieved using ``gating and in-painting'' to eliminate the influence of pre-ringdown data \cite{Zackay:2019kkv,Capano:2021etf}. 
We employ the gated-Gaussian likelihood implemented in the open-source PyCBC Inference library \cite{pycbcgithub,Biwer:2018osg}. This likelihood function applies gating and in-painting to gravitational wave data assuming the noise is a stationary Gaussian process. 

In all our analyses, a gate duration of two seconds is applied, with the gate ending at the selected start time of the ringdown phase. The power-spectral density (PSD) employed in the likelihood calculation is estimated from real detector data around the GW190521 event. This data has been publicly released by the Gravitational Wave Open Science Center \cite{LIGOScientific:2019lzm}. For all injections, the sky location (which determines the relative time, phase, and amplitude in each detector) of the ringdown template waveforms is fixed to the values obtained from the maximum likelihood result of Nitz \& Capano \cite{Nitz:2020mga}. To sample the parameter space during the inference process, we employ the \texttt{dynesty} nested sampler \cite{2020MNRAS.493.3132S}. This sampler efficiently explores the parameter space, provides reliable estimates of the posterior distribution and calculates the corresponding evidence. To speed convergence we numerically marginalize over the polarization of the gravitational wave using a discrete grid of 1000 points for each likelihood evaluation. This method has been shown to also improve the estimate of the evidence (and the Bayes factor, the ratio of evidences for two models)~\cite{Capano:2022zqm}.

In our study, we explore various signal models that encompass a range of angular and overtone modes, as defined by Eq.~\ref{linear_spectrum}. A complete list of all the models considered is given in Table~\ref{Table 1}. 
In all models, the complex frequencies of the modes are determined by the Kerr hypothesis, which provides predictions for the frequencies based on the final black hole's mass and spin (here calculated using the \texttt{pykerr} package~\cite{pykerr}). 

The priors for all parameters utilized in this study are provided in Table~\ref{table:prior}.
Priors for the amplitude parameters are identical to those in Capano et al. \cite{Capano:2021etf}.
Specifically, we select the amplitude of the $(2,2,1)$ mode to be within the range of [0, 5] times that of the $(2,2,0)$ mode. This choice is motivated by the numerical relativity fits presented in \cite{Giesler:2019uxc}. 
For the $(3,3,0)$ mode amplitude, we employ a prior that is [0, 0.5] times the amplitude of the $(2,2,0)$ mode. This choice is informed by the results of numerical simulations of binary black hole mergers described in Ref.~\cite{Borhanian:2019kxt}. Note that none of the simulated IMR signals studied in this paper had amplitudes outside this prior boundary.
For the $(2,1,0)$ and $(3,2,0)$ modes, we use the same prior distribution as for the $(3,3,0)$ mode. We note that Siegel et al. \cite{Siegel:2023lxl} find that the $(2,1,0)$ amplitude can exceed this range. Given that our objective in this study is to conduct a direct comparison of simulated signals with the real data from GW190521, specifically emphasizing the $(3,3,0)$ mode, the main conclusions of our work remain unaffected in light of this finding.


The maximum strain in the gravitational wave signal is expected to occur near the time of the merger. Using the numerical relativity (NR) surrogate waveform model NRSur7dq4, the LIGO Scientific and Virgo Collaborations (LVC) initially estimated the GPS time of the maximum strain in the Hanford detector to be $1242442967.4306^{+0.0067}_{-0.0106}$ (median $\pm 90\%$ credible interval) \cite{Varma:2019csw, LIGOScientific:2020ufj}. 
In the ringdown analysis, we consider discrete starting times for the analysis with $3$ms intervals. These starting times span $[-18, 24]$ms relative to the reference time $t_{\text{ref}}=1242442967.445$.\footnote{Please see next section for details.} This extends the range of $[-9, 24]$ms used in previous works \cite{Capano:2021etf, Capano:2022zqm}

\begin{table*}[t]
  \begin{center}
\begin{tabular}{||P{80pt}|p{280pt}|p{120pt}||}
\hline
\hline
{\bf Parameter} & \hfil {\bf Parameter description} & \hfil {\bf Uniform prior range} \\
\hline
\hline
$M_f$ & final black hole mass in the detector frame & [100,500] $M_\odot$\\
\hline
$\chi_f$ & final black hole spin &  [-0.99,0.99] \\
\hline
$\log_{10} A_{220}$ & base-10 logarithm of the amplitude of (2,2,0) & [-24,-19] \\
\hline
$A_{221/220}$ & ratio of amplitude between $(2,2,1)$ and $(2,2,0)$  &  [0,5] \\
\hline
$A_{330/220}$ & ratio of amplitudes between $(3,3,0)$ and $(2,2,0)$  &  [0,0.5] \\
\hline
$A_{210/220}$ & ratio of amplitudes between $(2,1,0)$ and $(2,2,0)$  &  [0,0.5] \\
\hline
$A_{320/220}$ & ratio of amplitudes between $(3,2,0)$ and $(2,2,0)$  &  [0,0.5] \\
\hline
$\phi_{220/221/330/210/320}$ & phase of (2,2,0)/(2,2,1)/(3,3,0)/(2,1,0)/(3,2,0) &  $[0,2\pi]$ \\
\hline
time-t$_{ref}$ & Discrete starting times (with 3 [ms] jumps) for the ringdown analysis & (-18, -15, -12, -9, -6, -3, 0, 3, 6, 9, 12, 15, 18, 21, 24) [ms] \\
\hline
\hline
\end{tabular}
\caption{The prior distributions of the sampling parameters for the models utilized in this study as defined by Eq.~\ref{linear_spectrum}.}
  \label{table:prior}
  \end{center}
\end{table*}

\section{Selection of simulated signals}
\label{sec:injectionsets}

We produce two distinct sets of simulated signals, (``injections'') denoted as set \emph{A} and set \emph{B}.
The signals are generated using the IMRPhenomTPHM waveform model, and their parameters are drawn from an IMR posterior found analysing GW190521 using this model \cite{Estelles:2021jnz}.
Each set consists of two subsets \emph{\signalinj{}} and \emph{\controlinj{}}.

Set \emph{A} is constructed such that its signals feature a $(3,3,0)$ ringdown mode with comparatively large amplitude. The signal parameters are drawn randomly from the posterior of \cite{Estelles:2021jnz}. We then select 100 signals with an amplitude of the IMR 33 mode at least 0.2 times as large as the amplitude of the 22 mode after merger. This selection is made by comparing the modes' amplitudes at a single point 
after the merger. Specifically, a time interval of $\delta t = 6 \text{ms} + t_{\text{ref}}^{L1} - t_{c}^{L1}$ is measured from the peak of the 22 IMR waveform. Here, $t_{\text{ref}}^{L1} = 1242442967.4243$ s (GPS time) denotes the estimated merger time of GW190521 at the Livingston detector ($t_{\text{ref}} = 1242442967.445$ s is the geocentric GPS reference time), and $t_{c}^{L1}$ stands for the coalescence time of the specific sample at the Livingston detector. In this comparison the amplitude of each IMR mode, e.g.\ $A_{22}$, is defined as $A_{22}=(h_{p,22}^2+h_{c,22}^{2})^{\frac{1}{2}}$, and equivalently for the $33$ mode. Here, $h_{p/c}$ represents the plus/cross polarization of the geocentric IMR waveform. Due to the differences between the IMR modes and the quasi-normal modes, this method does not directly constrain the amplitude of the $(3,3,0)$ QNM. However, this simple approach suffices to select samples where a significant $(3,3,0)$ mode is expected.


Through the selection, the signals in this set correspond to systems with higher mass-ratios, serving the purpose to test if GW190521 is an asymmetric system.
The parameter draws of set \emph{A} are identical to the 100 draws used in \cite{Capano:2022zqm}. 
There, we implemented an additional criterion that the signal-to-noise ratio (SNR) of the IMR 33 mode be at least 4, which restricted that study to a subset of set \emph{A}.

Set \emph{B} is chosen to consist of signals of nearly mass-symmetric systems from the GW190521 posterior. We expect this set to contain signals with comparatively small amplitudes for the subdominant ringdown modes \cite{Borhanian:2019kxt,JimenezForteza:2020cve}.
We construct this set by randomly selecting 9 points with mass ratios $m_1/m_2 < 2$ from the posterior distribution published in Estelles et al. \cite{Estelles:2021jnz}.
Adding the maximum likelihood point from the $m_1/m_2 < 2$ part of this posterior results in a total of 10 signals in this set.
This set provides a basis for comparison with results obtained from set \emph{A} to assess the significance of the asymmetric characteristics observed in GW190521.


For each set of signal parameters we generate two sets of signals.
The \emph{\signalinj{}} set is generated using all modes available in the approximant.
The \emph{\controlinj{}} set contains only the (IMR) 22 mode for this waveform. Both subsets of signals, \emph{\signalinj{}} and \emph{\controlinj{}}, use the same parameter samples drawn from the GW190521 posterior. The objective of the \emph{\controlinj{}} study is to examine whether subdominant modes are detected in the \emph{\signalinj{}} data and to investigate the start time of the late-time ringdown stage.


Similar to the findings presented in Nitz \& Capano \cite{Nitz:2020mga} with the NRSur7dq4 waveform model, Estelles et al.~\cite{Estelles:2021jnz} observed a bimodal posterior distribution in the component masses for GW190521. One mode favored nearly equal masses and another mode favored mass ratios of approximately 6:1. As it is depicted in Fig.~\ref{fig:imrcompare-time_mass_ratio_final_mass}, these bimodal features indicate the presence of two distinct configurations or scenarios for the binary black hole system associated with GW190521, characterized by different mass distributions between the two black holes. In our analysis, the injections labeled as set \emph{A} corresponds to the asymmetric configuration or mode, while the set \emph{B} captures the symmetric configuration of the GW190521 event.


Note that in this context, the notation $\ell m$ refers to the \emph{spherical} harmonics, which is the basis commonly used in inspiral-merger-ringdown (IMR) waveform models. The \emph{spherical} harmonics are distinct from the \emph{spheroidal} harmonics employed for quasi-normal modes (QNMs). While the QNM analysis uses the spheroidal harmonics to describe the ringdown phase, the IMR models employ the spherical harmonics to represent the overall behavior of the gravitational wave signal during the inspiral, merger, and ringdown stages. Therefore, our estimations of the Bayes factor in the absence of modes other than 22 can be considered as an upper bound on the underlying $(\ell,m,0)$ QNM mode. The presence of precession further complicates the picture, since precession causes modes with the same $\ell$ but different $m$ to mix together.
However, in this paper, we make the assumption that the late-stage behavior of the 22 mode in the IMR waveform closely resembles that of the $(2,2,0)+(2,2,1)$ QNMs. This assumption allows us to examine the behavior of the simulated signal injections specifically during the ringdown phase, while excluding the influence of QNMs other than $\ell=2$ and $m=2$.


To incorporate the injections into the analysis, both the \emph{\signalinj{}} and \emph{\controlinj{}} subsets are added to zero-noise data.
This means that the data consist of only the simulated signals, while the PSD used in the analysis is estimated from data around the event GW190521.
These injections are inserted at random times surrounding the estimated merger time of GW190521. The process follows the same methodology as described in Capano et al. \cite{Capano:2022zqm}.
There, an offset time $t_{\rm offset}$ is drawn uniformly from the range $\pm[4, 20]$ s. This offset time is then added to the coalescence time $t_c$, which is drawn from the relevant posterior distribution for each injection. The purpose of this offset is to introduce variability in the analysis and prevent any contamination from the original GW190521 data. A gap of $\pm4$ s around GW190521 is maintained to ensure the separation between the injections and the original event. Even though the presence of this gap is not essential when using zero-noise data, we here maintain the same methodology as used in Capano et al. for consistency \cite{Capano:2022zqm}.

The ringdown analysis is then performed on a grid of times surrounding each injection. The grid, used for the validation of the Kerr Bayes factor, spans the range $[-18,24]$ ms. This range is wider than the previous range used in Capano et al. \cite{Capano:2022zqm, Capano:2021etf}, which was $[-9,24]$ ms.

Similar to the analysis conducted in Capano et al. \cite{Capano:2021etf}, we establish a reference time, denoted as $t^{\rm inj}_{\rm ref}$, for each injection to construct the grid of times used in the ringdown analyses. For each injection, the reference time is defined as $t^{\rm inj}_{\rm ref} = t_{\rm ref} + t_{\rm offset}$, where $t_{\rm ref} = 1242442967.445$ GPS seconds represents the estimated geocentric merger time of GW190521 and $t_{\rm offset}$ is a randomly chosen offset unique to each injection. The reference time $t_{\rm ref}$ is obtained from the maximum likelihood parameters derived from the NRSurrogate analysis conducted in Nitz \& Capano \cite{Nitz:2020mga}. Note that $t^{\rm inj}_{\rm ref}$ is distinct from the injection's coalescence time $t^{\rm inj}_{c}$.

The ringdown analysis requires the identification of the time window during which the QNMs are observable. The QNM model is not valid at early times, such as before merger, where nonlinear components may be significant. However, if the analysis is conducted too long after the merger, the signal will have damped away to the extent that only the dominant mode remains observable. 

In the methodology presented in \cite{Capano:2022zqm, Capano:2021etf}, we address this challenge through the use of Bayes factors. 
A key objective of Bayesian inference is to derive a posterior probability distribution $p(\theta |d,M)$ for a set of model parameters $\theta$, given data $d$ and a model $M$ \cite{Thrane_2019}. Here, $d$ corresponds to the data from a (gravitational-wave) measurement, and the model $M$ describes the behaviour of noise and signal expected to produce the data. By applying Bayes' theorem, the posterior probability distribution is expressed as
\begin{equation}
p(\theta |d,M) = \frac{\mathcal{L}(d|\theta ,M) \pi(\theta |M)}{\mathcal{Z}(d|M)}.
\end{equation}
Here, $\mathcal{L}(d|\theta ,M)$ is the likelihood function of the data given the parameters $\theta$, which represents the probability of the detectors measuring data $d$ given a model hypothesis $M$ with source properties $\theta$. $\pi(\theta |M)$ is the prior distribution for $\theta$, encoding the prior knowledge about the parameters before studying the data $d$.
$\mathcal{Z}$ is a normalization factor called the “evidence” and quantifies how well the data aligns with the hypothesis,
\begin{eqnarray}
\mathcal{Z}(d|M) \equiv \int{d\theta \mathcal{L}(d|\theta ,M)\pi(\theta |M)}.
\end{eqnarray}

These provide a means to quantify the evidence that the data contain observable modes $\vec{X}={(2,2,0), ...}$ at a specific time $t-t_{\rm ref}$. By calculating the evidence $\mathcal{Z}_{\vec{X}}(t)$ and comparing it to the evidence for the $(2,2,0)$-only model ($\mathcal{Z}_{220}$) at the same time, we obtain the Bayes factor for the model. If model $\vec{X}$ and the $(2,2,0)$ model have equal prior weight (meaning that \textit{a priori} the two models are considered to be equally valid) then the Bayes factor gives the odds ratio for model $\vec{X}$ being true relative to the $(2, 2, 0)$-only model.

The $(2,2,0)$-only model is not a good representation of the signal at the merger \cite{Giesler:2019uxc}. Therefore, if we find a large value for $\mathcal{Z}_{\vec{X}}/\mathcal{Z}_{220}$ at a particular time, it is unclear whether this is due to the $\vec{X}$ modes being a good fit for the signal or if it is simply because the $(2,2,0)$-only model is inadequate at that time. In other words, the ratio $\mathcal{Z}_{\vec{X}}/\mathcal{Z}_{220}$ only indicates whether the $\vec{X}$ modes provide a better fit than the $(2,2,0)$-only model, not necessarily whether the $\vec{X}$ modes are truly observable. This issue becomes more pronounced as we approach the merger.

To address this concern, we leverage the observation from Ref.~\cite{Giesler:2019uxc} that including overtones of the dominant mode improves the fit to the signal close to or at the merger compared to the $(2,2,0)$-only model. We modify the Bayes factor as follows:
\begin{equation}
\label{eqn:bayes_factor}
\bayesfac(\vec{X}, t) \equiv \frac{\mathcal{Z}_{\vec{X}}(t)}{\max\{\mathcal{Z}_{220}, \mathcal{Z}_{220+221}\}}
\end{equation}
for all models $\vec{X}\neq (2,2,0)+(2,2,1)$. For the $(2,2,0)+(2,2,1)$ model, we simply use $\bayesfac=\mathcal{Z}_{220+221}/\mathcal{Z}_{220}$. This modification allows us to identify the most likely observable modes and determine the time at which they are most observable. By considering the maximum evidence between the $(2,2,0)$-only model and the model including an additional overtone, we account for the limitations of the $(2,2,0)$-only model and obtain a more robust assessment of the observability of the different modes at various times.

When applying this methodology to GW190521, the Bayes factor \bayesfac $(220+330)$ peaks at \eventbftime~with a value of \eventbf~\cite{Capano:2021etf,Capano:2022zqm}. Fig. \ref{fig:gw190521} shows the result of this methodology. This indicates that the $(2,2,0)+(3,3,0)$ model is approximately \approxebf{} times more likely to be true than the $(2,2,0)$-only model. We see that in Fig. \ref{fig:bfvsq} this peak time and value of bayes factor is consistent with what we obtain from GW190521.

We now apply this analysis to our injection sets.
Similar to the analysis of GW190521, we conduct the analysis on a grid of times spanning $t_{\rm ref} + [-18, 24]$ ms. However, to reduce computational costs given the large number of analyses, we sample the grid at intervals of $3$ms instead of the $1$ms intervals used in the previous work by Capano et al. \cite{Capano:2021etf}. The results of this analysis are shown in Figs. \ref{fig:QNMsfor100injections}, \ref{fig:bfvstime}, and \ref{fig:bfvstimeqlt2}.

\section{Results and Discussions}
\label{sec:conclusions}

In this paper, we investigate the nature of the GW190521 event by analyzing two sets of injections labeled as \emph{A} and \emph{B}, which are generated from the posterior samples of an IMR analysis of GW190521. Our analysis reveals that this event does not conform to a symmetric system. This is seen in the qualitative behavior of the time-dependent Bayes factors in favor of a subdominant mode. This finding further reinforces the previous observations of a bimodal posterior distribution in the component masses \cite{Nitz:2020mga,Estelles:2021jnz} (high mass ratio preference in Fig. \ref{fig:imrcompare-time_mass_ratio_final_mass}) using NRSur7dq4 and IMRPhenomTPHM waveforms. Figs. \ref{fig:QNMsfor100injections} and \ref{fig:bfvstime} demonstrate that as the mass ratio increases in the \emph{\signalinj} subset, the pattern of multiple modes in our study becomes more similar to the pattern observed in GW190521 (shown in Fig.~\ref{fig:gw190521}). However, when we perform the same analysis on the \emph{\controlinj} injections, which only contain the 22 mode, no resemblance to the observed pattern is found.

Remarkably, in Figs.~\ref{fig:QNMsfor100injections} and \ref{fig:bfvstime} for the high mass-ratio injections of the set \emph{A} \emph{\signalinj} subset, the peak Bayes factor corresponding to the (3,3,0) mode is observed to be located near the peak of GW190521 shown in Fig.~\ref{fig:gw190521}. The location of the peak is shown more clearly in the colorbar of Fig. \ref{fig:bfvsq} when considering the entire set A of injections. As the mass ratio increases, a distinct difference is observed when comparing the \emph{\signalinj} subset to the \emph{\controlinj} subset in both sets \emph{A} and \emph{B}. Specifically, in Figs. \ref{fig:QNMsfor100injections} and \ref{fig:bfvstime}, where high mass-ratio systems are present, there is no resemblance between the two subsets. However, the pattern in these figures closely matches that of GW190521 in Fig. \ref{fig:gw190521}.

In Fig. \ref{fig:bfvsq}, it is evident that the \emph{\signalinj} subset is shifted towards the Bayes factor value of GW190521 compared to the \emph{\controlinj} subset. This indicates that the inclusion of the (3,3,0) mode in the simulated signals for high mass ratio injections results in a similar Bayes factor value and pattern as observed for GW190521 in Fig. \ref{fig:gw190521}. 
For the low mass ratio injections in set B, both the \emph{\signalinj} and \emph{\controlinj} subsets exhibit no Bayes factor values close to that of GW190521. This observation is further supported by Fig. \ref{fig:bfvstimeqlt2}, where the observed patterns do not resemble that for GW190521 in Fig. \ref{fig:gw190521}. These findings raise questions regarding the symmetry of the GW190521 event, despite the injections being drawn from a part of the posterior distribution for GW190521. In both sets A and B, as we move to low mass ratios, the \emph{\signalinj} and \emph{\controlinj} subsets exhibit similar patterns, distinct from the observed pattern in GW190521 (Fig. \ref{fig:gw190521}).

In our analysis of simulated signals, we observe that the $(3,3,0)$ mode exhibits a more prominent peak compared to other combinations of modes studied in this paper. The presence of a peak in Bayes factor around time-t$_{ref} \simeq$ 6 ms for the \emph{\signalinj} subset, as opposed to its absence for the \emph{\controlinj} subset, suggests that at this point the linear (3,3,0) QNM is matching the full nonlinear 33 mode of the IMRPhenomTPHM model (see Fig.~\ref{fig:QNMsfor100injections}). This finding is further supported by this peak's absence for symmetric systems, in both subsets \emph{\signalinj} and \emph{\controlinj}. Therefore, this observation highlights the specific region in the ringdown phase of the GW190521 binary black hole merger event where spectroscopy can be effectively conducted. 

Furthermore, we observed that the values of the Bayes factors for high mass ratio systems in the simulated signals align with the results obtained from GW190521.


Generally, the Bayes factor serves to compare only two models, $\alpha$ and $\beta$. However, it's crucial to understand that a preference for model $\beta$ over model $\alpha$ does not necessarily imply that model $\beta$ is the accurate or true model. This is because both model $\alpha$ and model $\beta$ might not be appropriate models, and hence, a preference for model $\beta$ over model $\alpha$ does not validate model $\beta$ as a suitable representation. Therefore, it is important to acknowledge that while higher-order modes can seemingly fit the pre-merger regime in Figs.~\ref{fig:gw190521} and \ref{fig:QNMsfor100injections}, such a fit alone does not imply physical validity. The comparison of Bayes factors for the $(2,2,0)$ and $(2,2,0)+(2,2,1)$ models highlights that an improved fit is achieved with a higher number of modes, even though these models do not align well with the pre-merger time. Essentially, within the non-physical region, a greater number of modes facilitates a superior fit to the data.
Additionally, noise variations will introduce uncertainty in the values of the Bayes factors, as detailed in \cite{Capano:2022zqm}. Here, we only consider injections without  noise, approximating the behaviour expected when averaging over many instances of random noise.

In this analysis, since we employed injections without noise and the implications from black hole spectroscopy indicate that the late stage of black hole ringdown is accurately described by QNMs, we can confidently affirm the reliability of identifying the (3,3,0) mode using this method at later times.

In future work, we plan to study the recovery of intrinsic binary parameters from simulated signals through black hole spectroscopy.

As discussed in Section \ref{Introduction}, the NRSur7dq4 approximant is calibrated using NR simulations with mass ratios only up to $4$ and valid only up to mass ratios of $6$. Therefore, for achieving more precise results, more accurate waveforms are essential.

In conclusion, our findings strongly suggest that GW190521 is an asymmetric system that exhibits the presence of subdominant modes. While a more comprehensive analysis with a larger number of simulated signals is needed to obtain more precise evidence and uncertainties, our initial results indicate that GW190521 exhibits similarities to signals with mass ratios $\gtrapprox 8.5$. Moreover, our simulation results provide further support for the presence of the $(3,3,0)$ quasi-normal mode in the ringdown phase of GW190521.

\section*{Acknowledgments}
We thank Juan Calderón Bustillo, Gregorio Carullo and Harrison Siegel for useful comments.
J.\ A.\ was supported by ROMFORSK grant Project No.\ 302640. AHN acknowledges support from NSF grant PHY-2309240. We thank Atlas Computational Cluster team at the Albert Einstein Institute in Hanover for assistance. This research has made use of data obtained from the Gravitational Wave Open Science Center (https://www.gw-openscience.org/ ), a service of LIGO Laboratory, the LIGO Scientific Collaboration and the Virgo Collaboration. LIGO Laboratory and Advanced LIGO are funded by the United States National Science Foundation (NSF) who also gratefully acknowledge the Science and Technology Facilities Council (STFC) of the United Kingdom, the Max-Planck-Society (MPS), and the State of Niedersachsen/Germany for support of the construction of Advanced LIGO and construction and operation of the GEO600 detector. Additional support for Advanced LIGO was provided by the Australian Research Council. Virgo is funded, through the European Gravitational Observatory (EGO), by the French Centre National de Recherche Scientifique (CNRS), the Italian Istituto Nazionale di Fisica Nucleare (INFN) and the Dutch Nikhef, with contributions by institutions from Belgium, Germany, Greece, Hungary, Ireland, Japan, Monaco, Poland, Portugal, Spain. S.\ K.\ acknowledges support from the Villum Investigator program supported by VILLUM FONDEN (grant no. 37766)  and the DNRF Chair, by the Danish Research Foundation.

\bibliography{main.bib}

\begin{thebibliography}{69}%
\makeatletter
\providecommand \@ifxundefined [1]{%
 \@ifx{#1\undefined}
}%
\providecommand \@ifnum [1]{%
 \ifnum #1\expandafter \@firstoftwo
 \else \expandafter \@secondoftwo
 \fi
}%
\providecommand \@ifx [1]{%
 \ifx #1\expandafter \@firstoftwo
 \else \expandafter \@secondoftwo
 \fi
}%
\providecommand \natexlab [1]{#1}%
\providecommand \enquote  [1]{``#1''}%
\providecommand \bibnamefont  [1]{#1}%
\providecommand \bibfnamefont [1]{#1}%
\providecommand \citenamefont [1]{#1}%
\providecommand \href@noop [0]{\@secondoftwo}%
\providecommand \href [0]{\begingroup \@sanitize@url \@href}%
\providecommand \@href[1]{\@@startlink{#1}\@@href}%
\providecommand \@@href[1]{\endgroup#1\@@endlink}%
\providecommand \@sanitize@url [0]{\catcode `\\12\catcode `\$12\catcode
  `\&12\catcode `\#12\catcode `\^12\catcode `\_12\catcode `\%12\relax}%
\providecommand \@@startlink[1]{}%
\providecommand \@@endlink[0]{}%
\providecommand \url  [0]{\begingroup\@sanitize@url \@url }%
\providecommand \@url [1]{\endgroup\@href {#1}{\urlprefix }}%
\providecommand \urlprefix  [0]{URL }%
\providecommand \Eprint [0]{\href }%
\providecommand \doibase [0]{http://dx.doi.org/}%
\providecommand \selectlanguage [0]{\@gobble}%
\providecommand \bibinfo  [0]{\@secondoftwo}%
\providecommand \bibfield  [0]{\@secondoftwo}%
\providecommand \translation [1]{[#1]}%
\providecommand \BibitemOpen [0]{}%
\providecommand \bibitemStop [0]{}%
\providecommand \bibitemNoStop [0]{.\EOS\space}%
\providecommand \EOS [0]{\spacefactor3000\relax}%
\providecommand \BibitemShut  [1]{\csname bibitem#1\endcsname}%
\let\auto@bib@innerbib\@empty
\bibitem [{\citenamefont
  {Vishveshwara}(1970{\natexlab{a}})}]{Vishveshwara:1970cc}%
  \BibitemOpen
  \bibfield  {author} {\bibinfo {author} {\bibfnamefont {C.~V.}\ \bibnamefont
  {Vishveshwara}},\ }\href {\doibase 10.1103/PhysRevD.1.2870} {\bibfield
  {journal} {\bibinfo  {journal} {Phys. Rev. D}\ }\textbf {\bibinfo {volume}
  {1}},\ \bibinfo {pages} {2870} (\bibinfo {year}
  {1970}{\natexlab{a}})}\BibitemShut {NoStop}%
\bibitem [{\citenamefont
  {Vishveshwara}(1970{\natexlab{b}})}]{Vishveshwara:1970zz}%
  \BibitemOpen
  \bibfield  {author} {\bibinfo {author} {\bibfnamefont {C.~V.}\ \bibnamefont
  {Vishveshwara}},\ }\href {\doibase 10.1038/227936a0} {\bibfield  {journal}
  {\bibinfo  {journal} {Nature}\ }\textbf {\bibinfo {volume} {227}},\ \bibinfo
  {pages} {936} (\bibinfo {year} {1970}{\natexlab{b}})}\BibitemShut {NoStop}%
\bibitem [{\citenamefont {Chandrasekhar}\ and\ \citenamefont
  {Detweiler}(1975)}]{Chandrasekhar:1975zza}%
  \BibitemOpen
  \bibfield  {author} {\bibinfo {author} {\bibfnamefont {S.}~\bibnamefont
  {Chandrasekhar}}\ and\ \bibinfo {author} {\bibfnamefont {S.~L.}\ \bibnamefont
  {Detweiler}},\ }\href {\doibase 10.1098/rspa.1975.0112} {\bibfield  {journal}
  {\bibinfo  {journal} {Proc. Roy. Soc. Lond. A}\ }\textbf {\bibinfo {volume}
  {344}},\ \bibinfo {pages} {441} (\bibinfo {year} {1975})}\BibitemShut
  {NoStop}%
\bibitem [{\citenamefont {Wade}\ \emph {et~al.}(2013)\citenamefont {Wade},
  \citenamefont {Creighton}, \citenamefont {Ochsner},\ and\ \citenamefont
  {Nielsen}}]{Wade:2013hoa}%
  \BibitemOpen
  \bibfield  {author} {\bibinfo {author} {\bibfnamefont {M.}~\bibnamefont
  {Wade}}, \bibinfo {author} {\bibfnamefont {J.~D.~E.}\ \bibnamefont
  {Creighton}}, \bibinfo {author} {\bibfnamefont {E.}~\bibnamefont {Ochsner}},
  \ and\ \bibinfo {author} {\bibfnamefont {A.~B.}\ \bibnamefont {Nielsen}},\
  }\href {\doibase 10.1103/PhysRevD.88.083002} {\bibfield  {journal} {\bibinfo
  {journal} {Phys. Rev. D}\ }\textbf {\bibinfo {volume} {88}},\ \bibinfo
  {pages} {083002} (\bibinfo {year} {2013})},\ \Eprint
  {http://arxiv.org/abs/1306.3901} {arXiv:1306.3901 [gr-qc]} \BibitemShut
  {NoStop}%
\bibitem [{\citenamefont {Jim\'enez~Forteza}\ \emph {et~al.}(2020)\citenamefont
  {Jim\'enez~Forteza}, \citenamefont {Bhagwat}, \citenamefont {Pani},\ and\
  \citenamefont {Ferrari}}]{JimenezForteza:2020cve}%
  \BibitemOpen
  \bibfield  {author} {\bibinfo {author} {\bibfnamefont {X.}~\bibnamefont
  {Jim\'enez~Forteza}}, \bibinfo {author} {\bibfnamefont {S.}~\bibnamefont
  {Bhagwat}}, \bibinfo {author} {\bibfnamefont {P.}~\bibnamefont {Pani}}, \
  and\ \bibinfo {author} {\bibfnamefont {V.}~\bibnamefont {Ferrari}},\ }\href
  {\doibase 10.1103/PhysRevD.102.044053} {\bibfield  {journal} {\bibinfo
  {journal} {Phys. Rev. D}\ }\textbf {\bibinfo {volume} {102}},\ \bibinfo
  {pages} {044053} (\bibinfo {year} {2020})},\ \Eprint
  {http://arxiv.org/abs/2005.03260} {arXiv:2005.03260 [gr-qc]} \BibitemShut
  {NoStop}%
\bibitem [{\citenamefont {Borhanian}\ \emph {et~al.}(2020)\citenamefont
  {Borhanian}, \citenamefont {Arun}, \citenamefont {Pfeiffer},\ and\
  \citenamefont {Sathyaprakash}}]{Borhanian:2019kxt}%
  \BibitemOpen
  \bibfield  {author} {\bibinfo {author} {\bibfnamefont {S.}~\bibnamefont
  {Borhanian}}, \bibinfo {author} {\bibfnamefont {K.~G.}\ \bibnamefont {Arun}},
  \bibinfo {author} {\bibfnamefont {H.~P.}\ \bibnamefont {Pfeiffer}}, \ and\
  \bibinfo {author} {\bibfnamefont {B.~S.}\ \bibnamefont {Sathyaprakash}},\
  }\href {\doibase 10.1088/1361-6382/ab6a21} {\bibfield  {journal} {\bibinfo
  {journal} {Class. Quant. Grav.}\ }\textbf {\bibinfo {volume} {37}},\ \bibinfo
  {pages} {065006} (\bibinfo {year} {2020})},\ \Eprint
  {http://arxiv.org/abs/1901.08516} {arXiv:1901.08516 [gr-qc]} \BibitemShut
  {NoStop}%
\bibitem [{\citenamefont {Dreyer}\ \emph {et~al.}(2004)\citenamefont {Dreyer},
  \citenamefont {Kelly}, \citenamefont {Krishnan}, \citenamefont {Finn},
  \citenamefont {Garrison},\ and\ \citenamefont
  {Lopez-Aleman}}]{Dreyer:2003bv}%
  \BibitemOpen
  \bibfield  {author} {\bibinfo {author} {\bibfnamefont {O.}~\bibnamefont
  {Dreyer}}, \bibinfo {author} {\bibfnamefont {B.~J.}\ \bibnamefont {Kelly}},
  \bibinfo {author} {\bibfnamefont {B.}~\bibnamefont {Krishnan}}, \bibinfo
  {author} {\bibfnamefont {L.~S.}\ \bibnamefont {Finn}}, \bibinfo {author}
  {\bibfnamefont {D.}~\bibnamefont {Garrison}}, \ and\ \bibinfo {author}
  {\bibfnamefont {R.}~\bibnamefont {Lopez-Aleman}},\ }\href {\doibase
  10.1088/0264-9381/21/4/003} {\bibfield  {journal} {\bibinfo  {journal}
  {Class. Quant. Grav.}\ }\textbf {\bibinfo {volume} {21}},\ \bibinfo {pages}
  {787} (\bibinfo {year} {2004})},\ \Eprint
  {http://arxiv.org/abs/gr-qc/0309007} {arXiv:gr-qc/0309007} \BibitemShut
  {NoStop}%
\bibitem [{\citenamefont {Kamaretsos}\ \emph {et~al.}(2012)\citenamefont
  {Kamaretsos}, \citenamefont {Hannam}, \citenamefont {Husa},\ and\
  \citenamefont {Sathyaprakash}}]{Kamaretsos:2011um}%
  \BibitemOpen
  \bibfield  {author} {\bibinfo {author} {\bibfnamefont {I.}~\bibnamefont
  {Kamaretsos}}, \bibinfo {author} {\bibfnamefont {M.}~\bibnamefont {Hannam}},
  \bibinfo {author} {\bibfnamefont {S.}~\bibnamefont {Husa}}, \ and\ \bibinfo
  {author} {\bibfnamefont {B.~S.}\ \bibnamefont {Sathyaprakash}},\ }\href
  {\doibase 10.1103/PhysRevD.85.024018} {\bibfield  {journal} {\bibinfo
  {journal} {Phys. Rev. D}\ }\textbf {\bibinfo {volume} {85}},\ \bibinfo
  {pages} {024018} (\bibinfo {year} {2012})},\ \Eprint
  {http://arxiv.org/abs/1107.0854} {arXiv:1107.0854 [gr-qc]} \BibitemShut
  {NoStop}%
\bibitem [{\citenamefont {Bhagwat}\ \emph {et~al.}(2018)\citenamefont
  {Bhagwat}, \citenamefont {Okounkova}, \citenamefont {Ballmer}, \citenamefont
  {Brown}, \citenamefont {Giesler}, \citenamefont {Scheel},\ and\ \citenamefont
  {Teukolsky}}]{Bhagwat:2017tkm}%
  \BibitemOpen
  \bibfield  {author} {\bibinfo {author} {\bibfnamefont {S.}~\bibnamefont
  {Bhagwat}}, \bibinfo {author} {\bibfnamefont {M.}~\bibnamefont {Okounkova}},
  \bibinfo {author} {\bibfnamefont {S.~W.}\ \bibnamefont {Ballmer}}, \bibinfo
  {author} {\bibfnamefont {D.~A.}\ \bibnamefont {Brown}}, \bibinfo {author}
  {\bibfnamefont {M.}~\bibnamefont {Giesler}}, \bibinfo {author} {\bibfnamefont
  {M.~A.}\ \bibnamefont {Scheel}}, \ and\ \bibinfo {author} {\bibfnamefont
  {S.~A.}\ \bibnamefont {Teukolsky}},\ }\href {\doibase
  10.1103/PhysRevD.97.104065} {\bibfield  {journal} {\bibinfo  {journal} {Phys.
  Rev. D}\ }\textbf {\bibinfo {volume} {97}},\ \bibinfo {pages} {104065}
  (\bibinfo {year} {2018})},\ \Eprint {http://arxiv.org/abs/1711.00926}
  {arXiv:1711.00926 [gr-qc]} \BibitemShut {NoStop}%
\bibitem [{\citenamefont {Thrane}\ \emph {et~al.}(2017)\citenamefont {Thrane},
  \citenamefont {Lasky},\ and\ \citenamefont {Levin}}]{Thrane:2017lqn}%
  \BibitemOpen
  \bibfield  {author} {\bibinfo {author} {\bibfnamefont {E.}~\bibnamefont
  {Thrane}}, \bibinfo {author} {\bibfnamefont {P.~D.}\ \bibnamefont {Lasky}}, \
  and\ \bibinfo {author} {\bibfnamefont {Y.}~\bibnamefont {Levin}},\ }\href
  {\doibase 10.1103/PhysRevD.96.102004} {\bibfield  {journal} {\bibinfo
  {journal} {Phys. Rev. D}\ }\textbf {\bibinfo {volume} {96}},\ \bibinfo
  {pages} {102004} (\bibinfo {year} {2017})},\ \Eprint
  {http://arxiv.org/abs/1706.05152} {arXiv:1706.05152 [gr-qc]} \BibitemShut
  {NoStop}%
\bibitem [{\citenamefont {London}\ \emph {et~al.}(2014)\citenamefont {London},
  \citenamefont {Shoemaker},\ and\ \citenamefont {Healy}}]{London:2014cma}%
  \BibitemOpen
  \bibfield  {author} {\bibinfo {author} {\bibfnamefont {L.}~\bibnamefont
  {London}}, \bibinfo {author} {\bibfnamefont {D.}~\bibnamefont {Shoemaker}}, \
  and\ \bibinfo {author} {\bibfnamefont {J.}~\bibnamefont {Healy}},\ }\href
  {\doibase 10.1103/PhysRevD.90.124032} {\bibfield  {journal} {\bibinfo
  {journal} {Phys. Rev. D}\ }\textbf {\bibinfo {volume} {90}},\ \bibinfo
  {pages} {124032} (\bibinfo {year} {2014})},\ \bibinfo {note} {[Erratum:
  Phys.Rev.D 94, 069902 (2016)]},\ \Eprint {http://arxiv.org/abs/1404.3197}
  {arXiv:1404.3197 [gr-qc]} \BibitemShut {NoStop}%
\bibitem [{\citenamefont {Okounkova}(2020)}]{Okounkova:2020vwu}%
  \BibitemOpen
  \bibfield  {author} {\bibinfo {author} {\bibfnamefont {M.}~\bibnamefont
  {Okounkova}},\ }\href@noop {} {\  (\bibinfo {year} {2020})},\ \Eprint
  {http://arxiv.org/abs/2004.00671} {arXiv:2004.00671 [gr-qc]} \BibitemShut
  {NoStop}%
\bibitem [{\citenamefont {Mitman}\ \emph {et~al.}(2022)\citenamefont {Mitman}
  \emph {et~al.}}]{Mitman:2022qdl}%
  \BibitemOpen
  \bibfield  {author} {\bibinfo {author} {\bibfnamefont {K.}~\bibnamefont
  {Mitman}} \emph {et~al.},\ }\href@noop {} {\  (\bibinfo {year} {2022})},\
  \Eprint {http://arxiv.org/abs/2208.07380} {arXiv:2208.07380 [gr-qc]}
  \BibitemShut {NoStop}%
\bibitem [{\citenamefont {Lagos}\ and\ \citenamefont
  {Hui}(2022)}]{Lagos:2022otp}%
  \BibitemOpen
  \bibfield  {author} {\bibinfo {author} {\bibfnamefont {M.}~\bibnamefont
  {Lagos}}\ and\ \bibinfo {author} {\bibfnamefont {L.}~\bibnamefont {Hui}},\
  }\href@noop {} {\  (\bibinfo {year} {2022})},\ \Eprint
  {http://arxiv.org/abs/2208.07379} {arXiv:2208.07379 [gr-qc]} \BibitemShut
  {NoStop}%
\bibitem [{\citenamefont {Cheung}\ \emph {et~al.}(2022)\citenamefont {Cheung}
  \emph {et~al.}}]{Cheung:2022rbm}%
  \BibitemOpen
  \bibfield  {author} {\bibinfo {author} {\bibfnamefont {M.~H.-Y.}\
  \bibnamefont {Cheung}} \emph {et~al.},\ }\href@noop {} {\  (\bibinfo {year}
  {2022})},\ \Eprint {http://arxiv.org/abs/2208.07374} {arXiv:2208.07374
  [gr-qc]} \BibitemShut {NoStop}%
\bibitem [{\citenamefont {Jaramillo}\ \emph
  {et~al.}(2012{\natexlab{a}})\citenamefont {Jaramillo}, \citenamefont
  {Macedo}, \citenamefont {Moesta},\ and\ \citenamefont
  {Rezzolla}}]{Jaramillo:2012rr}%
  \BibitemOpen
  \bibfield  {author} {\bibinfo {author} {\bibfnamefont {J.~L.}\ \bibnamefont
  {Jaramillo}}, \bibinfo {author} {\bibfnamefont {R.~P.}\ \bibnamefont
  {Macedo}}, \bibinfo {author} {\bibfnamefont {P.}~\bibnamefont {Moesta}}, \
  and\ \bibinfo {author} {\bibfnamefont {L.}~\bibnamefont {Rezzolla}},\ }\href
  {\doibase 10.1063/1.4734411} {\bibfield  {journal} {\bibinfo  {journal} {AIP
  Conf. Proc.}\ }\textbf {\bibinfo {volume} {1458}},\ \bibinfo {pages} {158}
  (\bibinfo {year} {2012}{\natexlab{a}})},\ \Eprint
  {http://arxiv.org/abs/1205.3902} {arXiv:1205.3902 [gr-qc]} \BibitemShut
  {NoStop}%
\bibitem [{\citenamefont {Jaramillo}\ \emph
  {et~al.}(2012{\natexlab{b}})\citenamefont {Jaramillo}, \citenamefont
  {Panosso~Macedo}, \citenamefont {Moesta},\ and\ \citenamefont
  {Rezzolla}}]{Jaramillo:2011re}%
  \BibitemOpen
  \bibfield  {author} {\bibinfo {author} {\bibfnamefont {J.~L.}\ \bibnamefont
  {Jaramillo}}, \bibinfo {author} {\bibfnamefont {R.}~\bibnamefont
  {Panosso~Macedo}}, \bibinfo {author} {\bibfnamefont {P.}~\bibnamefont
  {Moesta}}, \ and\ \bibinfo {author} {\bibfnamefont {L.}~\bibnamefont
  {Rezzolla}},\ }\href {\doibase 10.1103/PhysRevD.85.084030} {\bibfield
  {journal} {\bibinfo  {journal} {Phys. Rev. D}\ }\textbf {\bibinfo {volume}
  {85}},\ \bibinfo {pages} {084030} (\bibinfo {year} {2012}{\natexlab{b}})},\
  \Eprint {http://arxiv.org/abs/1108.0060} {arXiv:1108.0060 [gr-qc]}
  \BibitemShut {NoStop}%
\bibitem [{\citenamefont {Prasad}\ \emph {et~al.}(2020)\citenamefont {Prasad},
  \citenamefont {Gupta}, \citenamefont {Bose}, \citenamefont {Krishnan},\ and\
  \citenamefont {Schnetter}}]{Prasad:2020xgr}%
  \BibitemOpen
  \bibfield  {author} {\bibinfo {author} {\bibfnamefont {V.}~\bibnamefont
  {Prasad}}, \bibinfo {author} {\bibfnamefont {A.}~\bibnamefont {Gupta}},
  \bibinfo {author} {\bibfnamefont {S.}~\bibnamefont {Bose}}, \bibinfo {author}
  {\bibfnamefont {B.}~\bibnamefont {Krishnan}}, \ and\ \bibinfo {author}
  {\bibfnamefont {E.}~\bibnamefont {Schnetter}},\ }\href {\doibase
  10.1103/PhysRevLett.125.121101} {\bibfield  {journal} {\bibinfo  {journal}
  {Phys. Rev. Lett.}\ }\textbf {\bibinfo {volume} {125}},\ \bibinfo {pages}
  {121101} (\bibinfo {year} {2020})},\ \Eprint
  {http://arxiv.org/abs/2003.06215} {arXiv:2003.06215 [gr-qc]} \BibitemShut
  {NoStop}%
\bibitem [{\citenamefont {Mourier}\ \emph {et~al.}(2021)\citenamefont
  {Mourier}, \citenamefont {Jim\'enez~Forteza}, \citenamefont {Pook-Kolb},
  \citenamefont {Krishnan},\ and\ \citenamefont {Schnetter}}]{Mourier:2020mwa}%
  \BibitemOpen
  \bibfield  {author} {\bibinfo {author} {\bibfnamefont {P.}~\bibnamefont
  {Mourier}}, \bibinfo {author} {\bibfnamefont {X.}~\bibnamefont
  {Jim\'enez~Forteza}}, \bibinfo {author} {\bibfnamefont {D.}~\bibnamefont
  {Pook-Kolb}}, \bibinfo {author} {\bibfnamefont {B.}~\bibnamefont {Krishnan}},
  \ and\ \bibinfo {author} {\bibfnamefont {E.}~\bibnamefont {Schnetter}},\
  }\href {\doibase 10.1103/PhysRevD.103.044054} {\bibfield  {journal} {\bibinfo
   {journal} {Phys. Rev. D}\ }\textbf {\bibinfo {volume} {103}},\ \bibinfo
  {pages} {044054} (\bibinfo {year} {2021})},\ \Eprint
  {http://arxiv.org/abs/2010.15186} {arXiv:2010.15186 [gr-qc]} \BibitemShut
  {NoStop}%
\bibitem [{\citenamefont {Forteza}\ and\ \citenamefont
  {Mourier}(2021)}]{Forteza:2021wfq}%
  \BibitemOpen
  \bibfield  {author} {\bibinfo {author} {\bibfnamefont {X.~J.}\ \bibnamefont
  {Forteza}}\ and\ \bibinfo {author} {\bibfnamefont {P.}~\bibnamefont
  {Mourier}},\ }\href {\doibase 10.1103/PhysRevD.104.124072} {\bibfield
  {journal} {\bibinfo  {journal} {Phys. Rev. D}\ }\textbf {\bibinfo {volume}
  {104}},\ \bibinfo {pages} {124072} (\bibinfo {year} {2021})},\ \Eprint
  {http://arxiv.org/abs/2107.11829} {arXiv:2107.11829 [gr-qc]} \BibitemShut
  {NoStop}%
\bibitem [{\citenamefont {Pook-Kolb}\ \emph {et~al.}(2020)\citenamefont
  {Pook-Kolb}, \citenamefont {Birnholtz}, \citenamefont {Jaramillo},
  \citenamefont {Krishnan},\ and\ \citenamefont
  {Schnetter}}]{Pook-Kolb:2020jlr}%
  \BibitemOpen
  \bibfield  {author} {\bibinfo {author} {\bibfnamefont {D.}~\bibnamefont
  {Pook-Kolb}}, \bibinfo {author} {\bibfnamefont {O.}~\bibnamefont
  {Birnholtz}}, \bibinfo {author} {\bibfnamefont {J.~L.}\ \bibnamefont
  {Jaramillo}}, \bibinfo {author} {\bibfnamefont {B.}~\bibnamefont {Krishnan}},
  \ and\ \bibinfo {author} {\bibfnamefont {E.}~\bibnamefont {Schnetter}},\
  }\href@noop {} {\  (\bibinfo {year} {2020})},\ \Eprint
  {http://arxiv.org/abs/2006.03940} {arXiv:2006.03940 [gr-qc]} \BibitemShut
  {NoStop}%
\bibitem [{\citenamefont {Gupta}\ \emph {et~al.}(2018)\citenamefont {Gupta},
  \citenamefont {Krishnan}, \citenamefont {Nielsen},\ and\ \citenamefont
  {Schnetter}}]{Gupta:2018znn}%
  \BibitemOpen
  \bibfield  {author} {\bibinfo {author} {\bibfnamefont {A.}~\bibnamefont
  {Gupta}}, \bibinfo {author} {\bibfnamefont {B.}~\bibnamefont {Krishnan}},
  \bibinfo {author} {\bibfnamefont {A.}~\bibnamefont {Nielsen}}, \ and\
  \bibinfo {author} {\bibfnamefont {E.}~\bibnamefont {Schnetter}},\ }\href
  {\doibase 10.1103/PhysRevD.97.084028} {\bibfield  {journal} {\bibinfo
  {journal} {Phys. Rev. D}\ }\textbf {\bibinfo {volume} {97}},\ \bibinfo
  {pages} {084028} (\bibinfo {year} {2018})},\ \Eprint
  {http://arxiv.org/abs/1801.07048} {arXiv:1801.07048 [gr-qc]} \BibitemShut
  {NoStop}%
\bibitem [{\citenamefont {Chen}\ \emph {et~al.}(2022)\citenamefont {Chen} \emph
  {et~al.}}]{Chen:2022dxt}%
  \BibitemOpen
  \bibfield  {author} {\bibinfo {author} {\bibfnamefont {Y.}~\bibnamefont
  {Chen}} \emph {et~al.},\ }\href@noop {} {\  (\bibinfo {year} {2022})},\
  \Eprint {http://arxiv.org/abs/2208.02965} {arXiv:2208.02965 [gr-qc]}
  \BibitemShut {NoStop}%
\bibitem [{\citenamefont {Khera}\ \emph {et~al.}(2023)\citenamefont {Khera},
  \citenamefont {Ribes~Metidieri}, \citenamefont {Bonga}, \citenamefont
  {Forteza}, \citenamefont {Krishnan}, \citenamefont {Poisson}, \citenamefont
  {Pook-Kolb}, \citenamefont {Schnetter},\ and\ \citenamefont
  {Yang}}]{Khera:2023lnc}%
  \BibitemOpen
  \bibfield  {author} {\bibinfo {author} {\bibfnamefont {N.}~\bibnamefont
  {Khera}}, \bibinfo {author} {\bibfnamefont {A.}~\bibnamefont
  {Ribes~Metidieri}}, \bibinfo {author} {\bibfnamefont {B.}~\bibnamefont
  {Bonga}}, \bibinfo {author} {\bibfnamefont {X.~J.}\ \bibnamefont {Forteza}},
  \bibinfo {author} {\bibfnamefont {B.}~\bibnamefont {Krishnan}}, \bibinfo
  {author} {\bibfnamefont {E.}~\bibnamefont {Poisson}}, \bibinfo {author}
  {\bibfnamefont {D.}~\bibnamefont {Pook-Kolb}}, \bibinfo {author}
  {\bibfnamefont {E.}~\bibnamefont {Schnetter}}, \ and\ \bibinfo {author}
  {\bibfnamefont {H.}~\bibnamefont {Yang}},\ }\href@noop {} {\  (\bibinfo
  {year} {2023})},\ \Eprint {http://arxiv.org/abs/2306.11142} {arXiv:2306.11142
  [gr-qc]} \BibitemShut {NoStop}%
\bibitem [{\citenamefont {Carullo}\ \emph {et~al.}(2018)\citenamefont {Carullo}
  \emph {et~al.}}]{Carullo:2018sfu}%
  \BibitemOpen
  \bibfield  {author} {\bibinfo {author} {\bibfnamefont {G.}~\bibnamefont
  {Carullo}} \emph {et~al.},\ }\href {\doibase 10.1103/PhysRevD.98.104020}
  {\bibfield  {journal} {\bibinfo  {journal} {Phys. Rev. D}\ }\textbf {\bibinfo
  {volume} {98}},\ \bibinfo {pages} {104020} (\bibinfo {year} {2018})},\
  \Eprint {http://arxiv.org/abs/1805.04760} {arXiv:1805.04760 [gr-qc]}
  \BibitemShut {NoStop}%
\bibitem [{\citenamefont {Carullo}\ \emph {et~al.}(2019)\citenamefont
  {Carullo}, \citenamefont {Del~Pozzo},\ and\ \citenamefont
  {Veitch}}]{Carullo:2019flw}%
  \BibitemOpen
  \bibfield  {author} {\bibinfo {author} {\bibfnamefont {G.}~\bibnamefont
  {Carullo}}, \bibinfo {author} {\bibfnamefont {W.}~\bibnamefont {Del~Pozzo}},
  \ and\ \bibinfo {author} {\bibfnamefont {J.}~\bibnamefont {Veitch}},\ }\href
  {\doibase 10.1103/PhysRevD.99.123029} {\bibfield  {journal} {\bibinfo
  {journal} {Phys. Rev. D}\ }\textbf {\bibinfo {volume} {99}},\ \bibinfo
  {pages} {123029} (\bibinfo {year} {2019})},\ \bibinfo {note} {[Erratum:
  Phys.Rev.D 100, 089903 (2019)]},\ \Eprint {http://arxiv.org/abs/1902.07527}
  {arXiv:1902.07527 [gr-qc]} \BibitemShut {NoStop}%
\bibitem [{\citenamefont {Cotesta}\ \emph {et~al.}(2022)\citenamefont
  {Cotesta}, \citenamefont {Carullo}, \citenamefont {Berti},\ and\
  \citenamefont {Cardoso}}]{Cotesta:2022pci}%
  \BibitemOpen
  \bibfield  {author} {\bibinfo {author} {\bibfnamefont {R.}~\bibnamefont
  {Cotesta}}, \bibinfo {author} {\bibfnamefont {G.}~\bibnamefont {Carullo}},
  \bibinfo {author} {\bibfnamefont {E.}~\bibnamefont {Berti}}, \ and\ \bibinfo
  {author} {\bibfnamefont {V.}~\bibnamefont {Cardoso}},\ }\href {\doibase
  10.1103/PhysRevLett.129.111102} {\bibfield  {journal} {\bibinfo  {journal}
  {Phys. Rev. Lett.}\ }\textbf {\bibinfo {volume} {129}},\ \bibinfo {pages}
  {111102} (\bibinfo {year} {2022})},\ \Eprint
  {http://arxiv.org/abs/2201.00822} {arXiv:2201.00822 [gr-qc]} \BibitemShut
  {NoStop}%
\bibitem [{\citenamefont {Cabero}\ \emph {et~al.}(2018)\citenamefont {Cabero},
  \citenamefont {Capano}, \citenamefont {Fischer-Birnholtz}, \citenamefont
  {Krishnan}, \citenamefont {Nielsen}, \citenamefont {Nitz},\ and\
  \citenamefont {Biwer}}]{Cabero:2017avf}%
  \BibitemOpen
  \bibfield  {author} {\bibinfo {author} {\bibfnamefont {M.}~\bibnamefont
  {Cabero}}, \bibinfo {author} {\bibfnamefont {C.~D.}\ \bibnamefont {Capano}},
  \bibinfo {author} {\bibfnamefont {O.}~\bibnamefont {Fischer-Birnholtz}},
  \bibinfo {author} {\bibfnamefont {B.}~\bibnamefont {Krishnan}}, \bibinfo
  {author} {\bibfnamefont {A.~B.}\ \bibnamefont {Nielsen}}, \bibinfo {author}
  {\bibfnamefont {A.~H.}\ \bibnamefont {Nitz}}, \ and\ \bibinfo {author}
  {\bibfnamefont {C.~M.}\ \bibnamefont {Biwer}},\ }\href {\doibase
  10.1103/PhysRevD.97.124069} {\bibfield  {journal} {\bibinfo  {journal} {Phys.
  Rev. D}\ }\textbf {\bibinfo {volume} {97}},\ \bibinfo {pages} {124069}
  (\bibinfo {year} {2018})},\ \Eprint {http://arxiv.org/abs/1711.09073}
  {arXiv:1711.09073 [gr-qc]} \BibitemShut {NoStop}%
\bibitem [{\citenamefont {Kastha}\ \emph {et~al.}(2022)\citenamefont {Kastha},
  \citenamefont {Capano}, \citenamefont {Westerweck}, \citenamefont {Cabero},
  \citenamefont {Krishnan},\ and\ \citenamefont {Nielsen}}]{Kastha:2021chr}%
  \BibitemOpen
  \bibfield  {author} {\bibinfo {author} {\bibfnamefont {S.}~\bibnamefont
  {Kastha}}, \bibinfo {author} {\bibfnamefont {C.~D.}\ \bibnamefont {Capano}},
  \bibinfo {author} {\bibfnamefont {J.}~\bibnamefont {Westerweck}}, \bibinfo
  {author} {\bibfnamefont {M.}~\bibnamefont {Cabero}}, \bibinfo {author}
  {\bibfnamefont {B.}~\bibnamefont {Krishnan}}, \ and\ \bibinfo {author}
  {\bibfnamefont {A.~B.}\ \bibnamefont {Nielsen}},\ }\href {\doibase
  10.1103/PhysRevD.105.064042} {\bibfield  {journal} {\bibinfo  {journal}
  {Phys. Rev. D}\ }\textbf {\bibinfo {volume} {105}},\ \bibinfo {pages}
  {064042} (\bibinfo {year} {2022})},\ \Eprint
  {http://arxiv.org/abs/2111.13664} {arXiv:2111.13664 [gr-qc]} \BibitemShut
  {NoStop}%
\bibitem [{\citenamefont {Berti}\ \emph {et~al.}(2016)\citenamefont {Berti},
  \citenamefont {Sesana}, \citenamefont {Barausse}, \citenamefont {Cardoso},\
  and\ \citenamefont {Belczynski}}]{Berti:2016lat}%
  \BibitemOpen
  \bibfield  {author} {\bibinfo {author} {\bibfnamefont {E.}~\bibnamefont
  {Berti}}, \bibinfo {author} {\bibfnamefont {A.}~\bibnamefont {Sesana}},
  \bibinfo {author} {\bibfnamefont {E.}~\bibnamefont {Barausse}}, \bibinfo
  {author} {\bibfnamefont {V.}~\bibnamefont {Cardoso}}, \ and\ \bibinfo
  {author} {\bibfnamefont {K.}~\bibnamefont {Belczynski}},\ }\href {\doibase
  10.1103/PhysRevLett.117.101102} {\bibfield  {journal} {\bibinfo  {journal}
  {Phys. Rev. Lett.}\ }\textbf {\bibinfo {volume} {117}},\ \bibinfo {pages}
  {101102} (\bibinfo {year} {2016})},\ \Eprint
  {http://arxiv.org/abs/1605.09286} {arXiv:1605.09286 [gr-qc]} \BibitemShut
  {NoStop}%
\bibitem [{\citenamefont {Cabero}\ \emph {et~al.}(2020)\citenamefont {Cabero},
  \citenamefont {Westerweck}, \citenamefont {Capano}, \citenamefont {Kumar},
  \citenamefont {Nielsen},\ and\ \citenamefont {Krishnan}}]{Cabero:2019zyt}%
  \BibitemOpen
  \bibfield  {author} {\bibinfo {author} {\bibfnamefont {M.}~\bibnamefont
  {Cabero}}, \bibinfo {author} {\bibfnamefont {J.}~\bibnamefont {Westerweck}},
  \bibinfo {author} {\bibfnamefont {C.~D.}\ \bibnamefont {Capano}}, \bibinfo
  {author} {\bibfnamefont {S.}~\bibnamefont {Kumar}}, \bibinfo {author}
  {\bibfnamefont {A.~B.}\ \bibnamefont {Nielsen}}, \ and\ \bibinfo {author}
  {\bibfnamefont {B.}~\bibnamefont {Krishnan}},\ }\href {\doibase
  10.1103/PhysRevD.101.064044} {\bibfield  {journal} {\bibinfo  {journal}
  {Phys. Rev. D}\ }\textbf {\bibinfo {volume} {101}},\ \bibinfo {pages}
  {064044} (\bibinfo {year} {2020})},\ \Eprint
  {http://arxiv.org/abs/1911.01361} {arXiv:1911.01361 [gr-qc]} \BibitemShut
  {NoStop}%
\bibitem [{\citenamefont {Isi}\ \emph {et~al.}(2019)\citenamefont {Isi},
  \citenamefont {Giesler}, \citenamefont {Farr}, \citenamefont {Scheel},\ and\
  \citenamefont {Teukolsky}}]{Isi:2019aib}%
  \BibitemOpen
  \bibfield  {author} {\bibinfo {author} {\bibfnamefont {M.}~\bibnamefont
  {Isi}}, \bibinfo {author} {\bibfnamefont {M.}~\bibnamefont {Giesler}},
  \bibinfo {author} {\bibfnamefont {W.~M.}\ \bibnamefont {Farr}}, \bibinfo
  {author} {\bibfnamefont {M.~A.}\ \bibnamefont {Scheel}}, \ and\ \bibinfo
  {author} {\bibfnamefont {S.~A.}\ \bibnamefont {Teukolsky}},\ }\href {\doibase
  10.1103/PhysRevLett.123.111102} {\bibfield  {journal} {\bibinfo  {journal}
  {Phys. Rev. Lett.}\ }\textbf {\bibinfo {volume} {123}},\ \bibinfo {pages}
  {111102} (\bibinfo {year} {2019})},\ \Eprint
  {http://arxiv.org/abs/1905.00869} {arXiv:1905.00869 [gr-qc]} \BibitemShut
  {NoStop}%
\bibitem [{\citenamefont {Giesler}\ \emph {et~al.}(2019)\citenamefont
  {Giesler}, \citenamefont {Isi}, \citenamefont {Scheel},\ and\ \citenamefont
  {Teukolsky}}]{Giesler:2019uxc}%
  \BibitemOpen
  \bibfield  {author} {\bibinfo {author} {\bibfnamefont {M.}~\bibnamefont
  {Giesler}}, \bibinfo {author} {\bibfnamefont {M.}~\bibnamefont {Isi}},
  \bibinfo {author} {\bibfnamefont {M.~A.}\ \bibnamefont {Scheel}}, \ and\
  \bibinfo {author} {\bibfnamefont {S.}~\bibnamefont {Teukolsky}},\ }\href
  {\doibase 10.1103/PhysRevX.9.041060} {\bibfield  {journal} {\bibinfo
  {journal} {Phys. Rev. X}\ }\textbf {\bibinfo {volume} {9}},\ \bibinfo {pages}
  {041060} (\bibinfo {year} {2019})},\ \Eprint
  {http://arxiv.org/abs/1903.08284} {arXiv:1903.08284 [gr-qc]} \BibitemShut
  {NoStop}%
\bibitem [{\citenamefont {Crisostomi}\ \emph {et~al.}(2023)\citenamefont
  {Crisostomi}, \citenamefont {Dey}, \citenamefont {Barausse},\ and\
  \citenamefont {Trotta}}]{Crisostomi:2023tle}%
  \BibitemOpen
  \bibfield  {author} {\bibinfo {author} {\bibfnamefont {M.}~\bibnamefont
  {Crisostomi}}, \bibinfo {author} {\bibfnamefont {K.}~\bibnamefont {Dey}},
  \bibinfo {author} {\bibfnamefont {E.}~\bibnamefont {Barausse}}, \ and\
  \bibinfo {author} {\bibfnamefont {R.}~\bibnamefont {Trotta}},\ }\href
  {\doibase 10.1103/PhysRevD.108.044029} {\bibfield  {journal} {\bibinfo
  {journal} {Phys. Rev. D}\ }\textbf {\bibinfo {volume} {108}},\ \bibinfo
  {pages} {044029} (\bibinfo {year} {2023})},\ \Eprint
  {http://arxiv.org/abs/2305.18528} {arXiv:2305.18528 [gr-qc]} \BibitemShut
  {NoStop}%
\bibitem [{\citenamefont {Isi}\ and\ \citenamefont {Farr}(2022)}]{Isi:2022mhy}%
  \BibitemOpen
  \bibfield  {author} {\bibinfo {author} {\bibfnamefont {M.}~\bibnamefont
  {Isi}}\ and\ \bibinfo {author} {\bibfnamefont {W.~M.}\ \bibnamefont {Farr}},\
  }\href@noop {} {\  (\bibinfo {year} {2022})},\ \Eprint
  {http://arxiv.org/abs/2202.02941} {arXiv:2202.02941 [gr-qc]} \BibitemShut
  {NoStop}%
\bibitem [{\citenamefont {Baibhav}\ \emph {et~al.}(2023)\citenamefont
  {Baibhav}, \citenamefont {Cheung}, \citenamefont {Berti}, \citenamefont
  {Cardoso}, \citenamefont {Carullo}, \citenamefont {Cotesta}, \citenamefont
  {Del~Pozzo},\ and\ \citenamefont {Duque}}]{Baibhav:2023clw}%
  \BibitemOpen
  \bibfield  {author} {\bibinfo {author} {\bibfnamefont {V.}~\bibnamefont
  {Baibhav}}, \bibinfo {author} {\bibfnamefont {M.~H.-Y.}\ \bibnamefont
  {Cheung}}, \bibinfo {author} {\bibfnamefont {E.}~\bibnamefont {Berti}},
  \bibinfo {author} {\bibfnamefont {V.}~\bibnamefont {Cardoso}}, \bibinfo
  {author} {\bibfnamefont {G.}~\bibnamefont {Carullo}}, \bibinfo {author}
  {\bibfnamefont {R.}~\bibnamefont {Cotesta}}, \bibinfo {author} {\bibfnamefont
  {W.}~\bibnamefont {Del~Pozzo}}, \ and\ \bibinfo {author} {\bibfnamefont
  {F.}~\bibnamefont {Duque}},\ }\href@noop {} {\  (\bibinfo {year} {2023})},\
  \Eprint {http://arxiv.org/abs/2302.03050} {arXiv:2302.03050 [gr-qc]}
  \BibitemShut {NoStop}%
\bibitem [{\citenamefont {Nollert}(1996)}]{Nollert:1996rf}%
  \BibitemOpen
  \bibfield  {author} {\bibinfo {author} {\bibfnamefont {H.-P.}\ \bibnamefont
  {Nollert}},\ }\href {\doibase 10.1103/PhysRevD.53.4397} {\bibfield  {journal}
  {\bibinfo  {journal} {Phys. Rev. D}\ }\textbf {\bibinfo {volume} {53}},\
  \bibinfo {pages} {4397} (\bibinfo {year} {1996})},\ \Eprint
  {http://arxiv.org/abs/gr-qc/9602032} {arXiv:gr-qc/9602032} \BibitemShut
  {NoStop}%
\bibitem [{\citenamefont {Nollert}\ and\ \citenamefont
  {Price}(1999)}]{Nollert:1998ys}%
  \BibitemOpen
  \bibfield  {author} {\bibinfo {author} {\bibfnamefont {H.-P.}\ \bibnamefont
  {Nollert}}\ and\ \bibinfo {author} {\bibfnamefont {R.~H.}\ \bibnamefont
  {Price}},\ }\href {\doibase 10.1063/1.532698} {\bibfield  {journal} {\bibinfo
   {journal} {J. Math. Phys.}\ }\textbf {\bibinfo {volume} {40}},\ \bibinfo
  {pages} {980} (\bibinfo {year} {1999})},\ \Eprint
  {http://arxiv.org/abs/gr-qc/9810074} {arXiv:gr-qc/9810074} \BibitemShut
  {NoStop}%
\bibitem [{\citenamefont {Jaramillo}\ \emph {et~al.}(2021)\citenamefont
  {Jaramillo}, \citenamefont {Panosso~Macedo},\ and\ \citenamefont
  {Al~Sheikh}}]{Jaramillo:2020tuu}%
  \BibitemOpen
  \bibfield  {author} {\bibinfo {author} {\bibfnamefont {J.~L.}\ \bibnamefont
  {Jaramillo}}, \bibinfo {author} {\bibfnamefont {R.}~\bibnamefont
  {Panosso~Macedo}}, \ and\ \bibinfo {author} {\bibfnamefont {L.}~\bibnamefont
  {Al~Sheikh}},\ }\href {\doibase 10.1103/PhysRevX.11.031003} {\bibfield
  {journal} {\bibinfo  {journal} {Phys. Rev. X}\ }\textbf {\bibinfo {volume}
  {11}},\ \bibinfo {pages} {031003} (\bibinfo {year} {2021})},\ \Eprint
  {http://arxiv.org/abs/2004.06434} {arXiv:2004.06434 [gr-qc]} \BibitemShut
  {NoStop}%
\bibitem [{\citenamefont {Jaramillo}\ \emph {et~al.}(2022)\citenamefont
  {Jaramillo}, \citenamefont {Panosso~Macedo},\ and\ \citenamefont
  {Sheikh}}]{Jaramillo:2021tmt}%
  \BibitemOpen
  \bibfield  {author} {\bibinfo {author} {\bibfnamefont {J.~L.}\ \bibnamefont
  {Jaramillo}}, \bibinfo {author} {\bibfnamefont {R.}~\bibnamefont
  {Panosso~Macedo}}, \ and\ \bibinfo {author} {\bibfnamefont {L.~A.}\
  \bibnamefont {Sheikh}},\ }\href {\doibase 10.1103/PhysRevLett.128.211102}
  {\bibfield  {journal} {\bibinfo  {journal} {Phys. Rev. Lett.}\ }\textbf
  {\bibinfo {volume} {128}},\ \bibinfo {pages} {211102} (\bibinfo {year}
  {2022})},\ \Eprint {http://arxiv.org/abs/2105.03451} {arXiv:2105.03451
  [gr-qc]} \BibitemShut {NoStop}%
\bibitem [{\citenamefont {Destounis}\ \emph {et~al.}(2021)\citenamefont
  {Destounis}, \citenamefont {Macedo}, \citenamefont {Berti}, \citenamefont
  {Cardoso},\ and\ \citenamefont {Jaramillo}}]{Destounis:2021lum}%
  \BibitemOpen
  \bibfield  {author} {\bibinfo {author} {\bibfnamefont {K.}~\bibnamefont
  {Destounis}}, \bibinfo {author} {\bibfnamefont {R.~P.}\ \bibnamefont
  {Macedo}}, \bibinfo {author} {\bibfnamefont {E.}~\bibnamefont {Berti}},
  \bibinfo {author} {\bibfnamefont {V.}~\bibnamefont {Cardoso}}, \ and\
  \bibinfo {author} {\bibfnamefont {J.~L.}\ \bibnamefont {Jaramillo}},\ }\href
  {\doibase 10.1103/PhysRevD.104.084091} {\bibfield  {journal} {\bibinfo
  {journal} {Phys. Rev. D}\ }\textbf {\bibinfo {volume} {104}},\ \bibinfo
  {pages} {084091} (\bibinfo {year} {2021})},\ \Eprint
  {http://arxiv.org/abs/2107.09673} {arXiv:2107.09673 [gr-qc]} \BibitemShut
  {NoStop}%
\bibitem [{\citenamefont {Abbott}\ \emph
  {et~al.}(2020{\natexlab{a}})\citenamefont {Abbott} \emph
  {et~al.}}]{LIGOScientific:2020iuh}%
  \BibitemOpen
  \bibfield  {author} {\bibinfo {author} {\bibfnamefont {R.}~\bibnamefont
  {Abbott}} \emph {et~al.} (\bibinfo {collaboration} {LIGO Scientific,
  Virgo}),\ }\href {\doibase 10.1103/PhysRevLett.125.101102} {\bibfield
  {journal} {\bibinfo  {journal} {Phys. Rev. Lett.}\ }\textbf {\bibinfo
  {volume} {125}},\ \bibinfo {pages} {101102} (\bibinfo {year}
  {2020}{\natexlab{a}})},\ \Eprint {http://arxiv.org/abs/2009.01075}
  {arXiv:2009.01075 [gr-qc]} \BibitemShut {NoStop}%
\bibitem [{\citenamefont {Bustillo}\ \emph {et~al.}(2021)\citenamefont
  {Bustillo}, \citenamefont {Sanchis-Gual}, \citenamefont {Torres-Forn\'e},
  \citenamefont {Font}, \citenamefont {Vajpeyi}, \citenamefont {Smith},
  \citenamefont {Herdeiro}, \citenamefont {Radu},\ and\ \citenamefont
  {Leong}}]{Bustillo:2020syj}%
  \BibitemOpen
  \bibfield  {author} {\bibinfo {author} {\bibfnamefont {J.~C.}\ \bibnamefont
  {Bustillo}}, \bibinfo {author} {\bibfnamefont {N.}~\bibnamefont
  {Sanchis-Gual}}, \bibinfo {author} {\bibfnamefont {A.}~\bibnamefont
  {Torres-Forn\'e}}, \bibinfo {author} {\bibfnamefont {J.~A.}\ \bibnamefont
  {Font}}, \bibinfo {author} {\bibfnamefont {A.}~\bibnamefont {Vajpeyi}},
  \bibinfo {author} {\bibfnamefont {R.}~\bibnamefont {Smith}}, \bibinfo
  {author} {\bibfnamefont {C.}~\bibnamefont {Herdeiro}}, \bibinfo {author}
  {\bibfnamefont {E.}~\bibnamefont {Radu}}, \ and\ \bibinfo {author}
  {\bibfnamefont {S.~H.~W.}\ \bibnamefont {Leong}},\ }\href {\doibase
  10.1103/PhysRevLett.126.081101} {\bibfield  {journal} {\bibinfo  {journal}
  {Phys. Rev. Lett.}\ }\textbf {\bibinfo {volume} {126}},\ \bibinfo {pages}
  {081101} (\bibinfo {year} {2021})},\ \Eprint
  {http://arxiv.org/abs/2009.05376} {arXiv:2009.05376 [gr-qc]} \BibitemShut
  {NoStop}%
\bibitem [{\citenamefont {Romero-Shaw}\ \emph {et~al.}(2020)\citenamefont
  {Romero-Shaw}, \citenamefont {Lasky}, \citenamefont {Thrane},\ and\
  \citenamefont {Bustillo}}]{Romero-Shaw:2020thy}%
  \BibitemOpen
  \bibfield  {author} {\bibinfo {author} {\bibfnamefont {I.~M.}\ \bibnamefont
  {Romero-Shaw}}, \bibinfo {author} {\bibfnamefont {P.~D.}\ \bibnamefont
  {Lasky}}, \bibinfo {author} {\bibfnamefont {E.}~\bibnamefont {Thrane}}, \
  and\ \bibinfo {author} {\bibfnamefont {J.~C.}\ \bibnamefont {Bustillo}},\
  }\href {\doibase 10.3847/2041-8213/abbe26} {\bibfield  {journal} {\bibinfo
  {journal} {Astrophys. J. Lett.}\ }\textbf {\bibinfo {volume} {903}},\
  \bibinfo {pages} {L5} (\bibinfo {year} {2020})},\ \Eprint
  {http://arxiv.org/abs/2009.04771} {arXiv:2009.04771 [astro-ph.HE]}
  \BibitemShut {NoStop}%
\bibitem [{\citenamefont {Gayathri}\ \emph {et~al.}(2022)\citenamefont
  {Gayathri}, \citenamefont {Healy}, \citenamefont {Lange}, \citenamefont
  {O'Brien}, \citenamefont {Szczepanczyk}, \citenamefont {Bartos},
  \citenamefont {Campanelli}, \citenamefont {Klimenko}, \citenamefont
  {Lousto},\ and\ \citenamefont {O'Shaughnessy}}]{Gayathri:2020coq}%
  \BibitemOpen
  \bibfield  {author} {\bibinfo {author} {\bibfnamefont {V.}~\bibnamefont
  {Gayathri}}, \bibinfo {author} {\bibfnamefont {J.}~\bibnamefont {Healy}},
  \bibinfo {author} {\bibfnamefont {J.}~\bibnamefont {Lange}}, \bibinfo
  {author} {\bibfnamefont {B.}~\bibnamefont {O'Brien}}, \bibinfo {author}
  {\bibfnamefont {M.}~\bibnamefont {Szczepanczyk}}, \bibinfo {author}
  {\bibfnamefont {I.}~\bibnamefont {Bartos}}, \bibinfo {author} {\bibfnamefont
  {M.}~\bibnamefont {Campanelli}}, \bibinfo {author} {\bibfnamefont
  {S.}~\bibnamefont {Klimenko}}, \bibinfo {author} {\bibfnamefont {C.~O.}\
  \bibnamefont {Lousto}}, \ and\ \bibinfo {author} {\bibfnamefont
  {R.}~\bibnamefont {O'Shaughnessy}},\ }\href {\doibase
  10.1038/s41550-021-01568-w} {\bibfield  {journal} {\bibinfo  {journal}
  {Nature Astron.}\ }\textbf {\bibinfo {volume} {6}},\ \bibinfo {pages} {344}
  (\bibinfo {year} {2022})},\ \Eprint {http://arxiv.org/abs/2009.05461}
  {arXiv:2009.05461 [astro-ph.HE]} \BibitemShut {NoStop}%
\bibitem [{\citenamefont {Abedi}\ \emph {et~al.}(2021)\citenamefont {Abedi},
  \citenamefont {Micchi},\ and\ \citenamefont {Afshordi}}]{Abedi:2021tti}%
  \BibitemOpen
  \bibfield  {author} {\bibinfo {author} {\bibfnamefont {J.}~\bibnamefont
  {Abedi}}, \bibinfo {author} {\bibfnamefont {L.~F.~L.}\ \bibnamefont
  {Micchi}}, \ and\ \bibinfo {author} {\bibfnamefont {N.}~\bibnamefont
  {Afshordi}},\ }\href@noop {} {\  (\bibinfo {year} {2021})},\ \Eprint
  {http://arxiv.org/abs/2201.00047} {arXiv:2201.00047 [gr-qc]} \BibitemShut
  {NoStop}%
\bibitem [{\citenamefont {Wang}\ \emph {et~al.}(2022)\citenamefont {Wang},
  \citenamefont {Brown}, \citenamefont {Shao},\ and\ \citenamefont
  {Zhao}}]{Wang:2021gqm}%
  \BibitemOpen
  \bibfield  {author} {\bibinfo {author} {\bibfnamefont {Y.-F.}\ \bibnamefont
  {Wang}}, \bibinfo {author} {\bibfnamefont {S.~M.}\ \bibnamefont {Brown}},
  \bibinfo {author} {\bibfnamefont {L.}~\bibnamefont {Shao}}, \ and\ \bibinfo
  {author} {\bibfnamefont {W.}~\bibnamefont {Zhao}},\ }\href {\doibase
  10.1103/PhysRevD.106.084005} {\bibfield  {journal} {\bibinfo  {journal}
  {Phys. Rev. D}\ }\textbf {\bibinfo {volume} {106}},\ \bibinfo {pages}
  {084005} (\bibinfo {year} {2022})},\ \Eprint
  {http://arxiv.org/abs/2109.09718} {arXiv:2109.09718 [astro-ph.HE]}
  \BibitemShut {NoStop}%
\bibitem [{\citenamefont {Gamba}\ \emph {et~al.}(2021)\citenamefont {Gamba},
  \citenamefont {Breschi}, \citenamefont {Carullo}, \citenamefont {Rettegno},
  \citenamefont {Albanesi}, \citenamefont {Bernuzzi},\ and\ \citenamefont
  {Nagar}}]{Gamba:2021gap}%
  \BibitemOpen
  \bibfield  {author} {\bibinfo {author} {\bibfnamefont {R.}~\bibnamefont
  {Gamba}}, \bibinfo {author} {\bibfnamefont {M.}~\bibnamefont {Breschi}},
  \bibinfo {author} {\bibfnamefont {G.}~\bibnamefont {Carullo}}, \bibinfo
  {author} {\bibfnamefont {P.}~\bibnamefont {Rettegno}}, \bibinfo {author}
  {\bibfnamefont {S.}~\bibnamefont {Albanesi}}, \bibinfo {author}
  {\bibfnamefont {S.}~\bibnamefont {Bernuzzi}}, \ and\ \bibinfo {author}
  {\bibfnamefont {A.}~\bibnamefont {Nagar}},\ }\href@noop {} {\  (\bibinfo
  {year} {2021})},\ \Eprint {http://arxiv.org/abs/2106.05575} {arXiv:2106.05575
  [gr-qc]} \BibitemShut {NoStop}%
\bibitem [{\citenamefont {Dall'Amico}\ \emph {et~al.}(2021)\citenamefont
  {Dall'Amico}, \citenamefont {Mapelli}, \citenamefont {Di~Carlo},
  \citenamefont {Bouffanais}, \citenamefont {Rastello}, \citenamefont
  {Santoliquido}, \citenamefont {Ballone},\ and\ \citenamefont
  {Sedda}}]{DallAmico:2021umv}%
  \BibitemOpen
  \bibfield  {author} {\bibinfo {author} {\bibfnamefont {M.}~\bibnamefont
  {Dall'Amico}}, \bibinfo {author} {\bibfnamefont {M.}~\bibnamefont {Mapelli}},
  \bibinfo {author} {\bibfnamefont {U.~N.}\ \bibnamefont {Di~Carlo}}, \bibinfo
  {author} {\bibfnamefont {Y.}~\bibnamefont {Bouffanais}}, \bibinfo {author}
  {\bibfnamefont {S.}~\bibnamefont {Rastello}}, \bibinfo {author}
  {\bibfnamefont {F.}~\bibnamefont {Santoliquido}}, \bibinfo {author}
  {\bibfnamefont {A.}~\bibnamefont {Ballone}}, \ and\ \bibinfo {author}
  {\bibfnamefont {M.~A.}\ \bibnamefont {Sedda}},\ }\href {\doibase
  10.1093/mnras/stab2783} {\bibfield  {journal} {\bibinfo  {journal} {Mon. Not.
  Roy. Astron. Soc.}\ }\textbf {\bibinfo {volume} {508}},\ \bibinfo {pages}
  {3045} (\bibinfo {year} {2021})},\ \Eprint {http://arxiv.org/abs/2105.12757}
  {arXiv:2105.12757 [astro-ph.HE]} \BibitemShut {NoStop}%
\bibitem [{\citenamefont {Shibata}\ \emph {et~al.}(2021)\citenamefont
  {Shibata}, \citenamefont {Kiuchi}, \citenamefont {Fujibayashi},\ and\
  \citenamefont {Sekiguchi}}]{Shibata:2021sau}%
  \BibitemOpen
  \bibfield  {author} {\bibinfo {author} {\bibfnamefont {M.}~\bibnamefont
  {Shibata}}, \bibinfo {author} {\bibfnamefont {K.}~\bibnamefont {Kiuchi}},
  \bibinfo {author} {\bibfnamefont {S.}~\bibnamefont {Fujibayashi}}, \ and\
  \bibinfo {author} {\bibfnamefont {Y.}~\bibnamefont {Sekiguchi}},\ }\href
  {\doibase 10.1103/PhysRevD.103.063037} {\bibfield  {journal} {\bibinfo
  {journal} {Phys. Rev. D}\ }\textbf {\bibinfo {volume} {103}},\ \bibinfo
  {pages} {063037} (\bibinfo {year} {2021})},\ \Eprint
  {http://arxiv.org/abs/2101.05440} {arXiv:2101.05440 [astro-ph.HE]}
  \BibitemShut {NoStop}%
\bibitem [{\citenamefont {Abedi}(2022)}]{Abedi:2022bph}%
  \BibitemOpen
  \bibfield  {author} {\bibinfo {author} {\bibfnamefont {J.}~\bibnamefont
  {Abedi}},\ }\href@noop {} {\  (\bibinfo {year} {2022})},\ \Eprint
  {http://arxiv.org/abs/2301.00025} {arXiv:2301.00025 [gr-qc]} \BibitemShut
  {NoStop}%
\bibitem [{\citenamefont {Calderon~Bustillo}\ \emph {et~al.}(2022)\citenamefont
  {Calderon~Bustillo}, \citenamefont {Sanchis-Gual}, \citenamefont {Leong},
  \citenamefont {Chandra}, \citenamefont {Torres-Forne}, \citenamefont {Font},
  \citenamefont {Herdeiro}, \citenamefont {Radu}, \citenamefont {Wong},\ and\
  \citenamefont {Li}}]{CalderonBustillo:2022cja}%
  \BibitemOpen
  \bibfield  {author} {\bibinfo {author} {\bibfnamefont {J.}~\bibnamefont
  {Calderon~Bustillo}}, \bibinfo {author} {\bibfnamefont {N.}~\bibnamefont
  {Sanchis-Gual}}, \bibinfo {author} {\bibfnamefont {S.~H.~W.}\ \bibnamefont
  {Leong}}, \bibinfo {author} {\bibfnamefont {K.}~\bibnamefont {Chandra}},
  \bibinfo {author} {\bibfnamefont {A.}~\bibnamefont {Torres-Forne}}, \bibinfo
  {author} {\bibfnamefont {J.~A.}\ \bibnamefont {Font}}, \bibinfo {author}
  {\bibfnamefont {C.}~\bibnamefont {Herdeiro}}, \bibinfo {author}
  {\bibfnamefont {E.}~\bibnamefont {Radu}}, \bibinfo {author} {\bibfnamefont
  {I.~C.~F.}\ \bibnamefont {Wong}}, \ and\ \bibinfo {author} {\bibfnamefont
  {T.~G.~F.}\ \bibnamefont {Li}},\ }\href@noop {} {\  (\bibinfo {year}
  {2022})},\ \Eprint {http://arxiv.org/abs/2206.02551} {arXiv:2206.02551
  [gr-qc]} \BibitemShut {NoStop}%
\bibitem [{\citenamefont {Abbott}\ \emph
  {et~al.}(2020{\natexlab{b}})\citenamefont {Abbott} \emph
  {et~al.}}]{LIGOScientific:2020ufj}%
  \BibitemOpen
  \bibfield  {author} {\bibinfo {author} {\bibfnamefont {R.}~\bibnamefont
  {Abbott}} \emph {et~al.} (\bibinfo {collaboration} {LIGO Scientific,
  Virgo}),\ }\href {\doibase 10.3847/2041-8213/aba493} {\bibfield  {journal}
  {\bibinfo  {journal} {Astrophys. J. Lett.}\ }\textbf {\bibinfo {volume}
  {900}},\ \bibinfo {pages} {L13} (\bibinfo {year} {2020}{\natexlab{b}})},\
  \Eprint {http://arxiv.org/abs/2009.01190} {arXiv:2009.01190 [astro-ph.HE]}
  \BibitemShut {NoStop}%
\bibitem [{\citenamefont {Abbott}\ \emph
  {et~al.}(2021{\natexlab{a}})\citenamefont {Abbott} \emph
  {et~al.}}]{LIGOScientific:2021djp}%
  \BibitemOpen
  \bibfield  {author} {\bibinfo {author} {\bibfnamefont {R.}~\bibnamefont
  {Abbott}} \emph {et~al.} (\bibinfo {collaboration} {LIGO Scientific, VIRGO,
  KAGRA}),\ }\href@noop {} {\  (\bibinfo {year} {2021}{\natexlab{a}})},\
  \Eprint {http://arxiv.org/abs/2111.03606} {arXiv:2111.03606 [gr-qc]}
  \BibitemShut {NoStop}%
\bibitem [{\citenamefont {Nitz}\ \emph
  {et~al.}(2021{\natexlab{a}})\citenamefont {Nitz}, \citenamefont {Kumar},
  \citenamefont {Wang}, \citenamefont {Kastha}, \citenamefont {Wu},
  \citenamefont {Sch\"afer}, \citenamefont {Dhurkunde},\ and\ \citenamefont
  {Capano}}]{Nitz:2021zwj}%
  \BibitemOpen
  \bibfield  {author} {\bibinfo {author} {\bibfnamefont {A.~H.}\ \bibnamefont
  {Nitz}}, \bibinfo {author} {\bibfnamefont {S.}~\bibnamefont {Kumar}},
  \bibinfo {author} {\bibfnamefont {Y.-F.}\ \bibnamefont {Wang}}, \bibinfo
  {author} {\bibfnamefont {S.}~\bibnamefont {Kastha}}, \bibinfo {author}
  {\bibfnamefont {S.}~\bibnamefont {Wu}}, \bibinfo {author} {\bibfnamefont
  {M.}~\bibnamefont {Sch\"afer}}, \bibinfo {author} {\bibfnamefont
  {R.}~\bibnamefont {Dhurkunde}}, \ and\ \bibinfo {author} {\bibfnamefont
  {C.~D.}\ \bibnamefont {Capano}},\ }\href@noop {} {\  (\bibinfo {year}
  {2021}{\natexlab{a}})},\ \Eprint {http://arxiv.org/abs/2112.06878}
  {arXiv:2112.06878 [astro-ph.HE]} \BibitemShut {NoStop}%
\bibitem [{\citenamefont {Capano}\ \emph {et~al.}(2021)\citenamefont {Capano},
  \citenamefont {Cabero}, \citenamefont {Westerweck}, \citenamefont {Abedi},
  \citenamefont {Kastha}, \citenamefont {Nitz}, \citenamefont {Nielsen},\ and\
  \citenamefont {Krishnan}}]{Capano:2021etf}%
  \BibitemOpen
  \bibfield  {author} {\bibinfo {author} {\bibfnamefont {C.~D.}\ \bibnamefont
  {Capano}}, \bibinfo {author} {\bibfnamefont {M.}~\bibnamefont {Cabero}},
  \bibinfo {author} {\bibfnamefont {J.}~\bibnamefont {Westerweck}}, \bibinfo
  {author} {\bibfnamefont {J.}~\bibnamefont {Abedi}}, \bibinfo {author}
  {\bibfnamefont {S.}~\bibnamefont {Kastha}}, \bibinfo {author} {\bibfnamefont
  {A.~H.}\ \bibnamefont {Nitz}}, \bibinfo {author} {\bibfnamefont {A.~B.}\
  \bibnamefont {Nielsen}}, \ and\ \bibinfo {author} {\bibfnamefont
  {B.}~\bibnamefont {Krishnan}},\ }\href@noop {} {\  (\bibinfo {year}
  {2021})},\ \Eprint {http://arxiv.org/abs/2105.05238} {arXiv:2105.05238
  [gr-qc]} \BibitemShut {NoStop}%
\bibitem [{\citenamefont {Capano}\ \emph {et~al.}(2022)\citenamefont {Capano},
  \citenamefont {Abedi}, \citenamefont {Kastha}, \citenamefont {Nitz},
  \citenamefont {Westerweck}, \citenamefont {Wang}, \citenamefont {Cabero},
  \citenamefont {Nielsen},\ and\ \citenamefont {Krishnan}}]{Capano:2022zqm}%
  \BibitemOpen
  \bibfield  {author} {\bibinfo {author} {\bibfnamefont {C.~D.}\ \bibnamefont
  {Capano}}, \bibinfo {author} {\bibfnamefont {J.}~\bibnamefont {Abedi}},
  \bibinfo {author} {\bibfnamefont {S.}~\bibnamefont {Kastha}}, \bibinfo
  {author} {\bibfnamefont {A.~H.}\ \bibnamefont {Nitz}}, \bibinfo {author}
  {\bibfnamefont {J.}~\bibnamefont {Westerweck}}, \bibinfo {author}
  {\bibfnamefont {Y.-F.}\ \bibnamefont {Wang}}, \bibinfo {author}
  {\bibfnamefont {M.}~\bibnamefont {Cabero}}, \bibinfo {author} {\bibfnamefont
  {A.~B.}\ \bibnamefont {Nielsen}}, \ and\ \bibinfo {author} {\bibfnamefont
  {B.}~\bibnamefont {Krishnan}},\ }\href@noop {} {\  (\bibinfo {year}
  {2022})},\ \Eprint {http://arxiv.org/abs/2209.00640} {arXiv:2209.00640
  [gr-qc]} \BibitemShut {NoStop}%
\bibitem [{\citenamefont {Siegel}\ \emph {et~al.}(2023)\citenamefont {Siegel},
  \citenamefont {Isi},\ and\ \citenamefont {Farr}}]{Siegel:2023lxl}%
  \BibitemOpen
  \bibfield  {author} {\bibinfo {author} {\bibfnamefont {H.}~\bibnamefont
  {Siegel}}, \bibinfo {author} {\bibfnamefont {M.}~\bibnamefont {Isi}}, \ and\
  \bibinfo {author} {\bibfnamefont {W.}~\bibnamefont {Farr}},\ }\href@noop {}
  {\  (\bibinfo {year} {2023})},\ \Eprint {http://arxiv.org/abs/2307.11975}
  {arXiv:2307.11975 [gr-qc]} \BibitemShut {NoStop}%
\bibitem [{\citenamefont {Varma}\ \emph {et~al.}(2019)\citenamefont {Varma},
  \citenamefont {Field}, \citenamefont {Scheel}, \citenamefont {Blackman},
  \citenamefont {Gerosa}, \citenamefont {Stein}, \citenamefont {Kidder},\ and\
  \citenamefont {Pfeiffer}}]{Varma:2019csw}%
  \BibitemOpen
  \bibfield  {author} {\bibinfo {author} {\bibfnamefont {V.}~\bibnamefont
  {Varma}}, \bibinfo {author} {\bibfnamefont {S.~E.}\ \bibnamefont {Field}},
  \bibinfo {author} {\bibfnamefont {M.~A.}\ \bibnamefont {Scheel}}, \bibinfo
  {author} {\bibfnamefont {J.}~\bibnamefont {Blackman}}, \bibinfo {author}
  {\bibfnamefont {D.}~\bibnamefont {Gerosa}}, \bibinfo {author} {\bibfnamefont
  {L.~C.}\ \bibnamefont {Stein}}, \bibinfo {author} {\bibfnamefont {L.~E.}\
  \bibnamefont {Kidder}}, \ and\ \bibinfo {author} {\bibfnamefont {H.~P.}\
  \bibnamefont {Pfeiffer}},\ }\href {\doibase 10.1103/PhysRevResearch.1.033015}
  {\bibfield  {journal} {\bibinfo  {journal} {Phys. Rev. Research.}\ }\textbf
  {\bibinfo {volume} {1}},\ \bibinfo {pages} {033015} (\bibinfo {year}
  {2019})},\ \Eprint {http://arxiv.org/abs/1905.09300} {arXiv:1905.09300
  [gr-qc]} \BibitemShut {NoStop}%
\bibitem [{\citenamefont {Nitz}\ and\ \citenamefont
  {Capano}(2021)}]{Nitz:2020mga}%
  \BibitemOpen
  \bibfield  {author} {\bibinfo {author} {\bibfnamefont {A.~H.}\ \bibnamefont
  {Nitz}}\ and\ \bibinfo {author} {\bibfnamefont {C.~D.}\ \bibnamefont
  {Capano}},\ }\href {\doibase 10.3847/2041-8213/abccc5} {\bibfield  {journal}
  {\bibinfo  {journal} {Astrophys. J. Lett.}\ }\textbf {\bibinfo {volume}
  {907}},\ \bibinfo {pages} {L9} (\bibinfo {year} {2021})},\ \Eprint
  {http://arxiv.org/abs/2010.12558} {arXiv:2010.12558 [astro-ph.HE]}
  \BibitemShut {NoStop}%
\bibitem [{\citenamefont {Estell\'es}\ \emph
  {et~al.}(2022{\natexlab{a}})\citenamefont {Estell\'es} \emph
  {et~al.}}]{Estelles:2021jnz}%
  \BibitemOpen
  \bibfield  {author} {\bibinfo {author} {\bibfnamefont {H.}~\bibnamefont
  {Estell\'es}} \emph {et~al.},\ }\href {\doibase 10.3847/1538-4357/ac33a0}
  {\bibfield  {journal} {\bibinfo  {journal} {Astrophys. J.}\ }\textbf
  {\bibinfo {volume} {924}},\ \bibinfo {pages} {79} (\bibinfo {year}
  {2022}{\natexlab{a}})},\ \Eprint {http://arxiv.org/abs/2105.06360}
  {arXiv:2105.06360 [gr-qc]} \BibitemShut {NoStop}%
\bibitem [{\citenamefont {Estell\'es}\ \emph
  {et~al.}(2022{\natexlab{b}})\citenamefont {Estell\'es}, \citenamefont
  {Colleoni}, \citenamefont {Garc\'\i{}a-Quir\'os}, \citenamefont {Husa},
  \citenamefont {Keitel}, \citenamefont {Mateu-Lucena}, \citenamefont
  {Planas},\ and\ \citenamefont {Ramos-Buades}}]{Estelles:2021gvs}%
  \BibitemOpen
  \bibfield  {author} {\bibinfo {author} {\bibfnamefont {H.}~\bibnamefont
  {Estell\'es}}, \bibinfo {author} {\bibfnamefont {M.}~\bibnamefont
  {Colleoni}}, \bibinfo {author} {\bibfnamefont {C.}~\bibnamefont
  {Garc\'\i{}a-Quir\'os}}, \bibinfo {author} {\bibfnamefont {S.}~\bibnamefont
  {Husa}}, \bibinfo {author} {\bibfnamefont {D.}~\bibnamefont {Keitel}},
  \bibinfo {author} {\bibfnamefont {M.}~\bibnamefont {Mateu-Lucena}}, \bibinfo
  {author} {\bibfnamefont {M.~d.~L.}\ \bibnamefont {Planas}}, \ and\ \bibinfo
  {author} {\bibfnamefont {A.}~\bibnamefont {Ramos-Buades}},\ }\href {\doibase
  10.1103/PhysRevD.105.084040} {\bibfield  {journal} {\bibinfo  {journal}
  {Phys. Rev. D}\ }\textbf {\bibinfo {volume} {105}},\ \bibinfo {pages}
  {084040} (\bibinfo {year} {2022}{\natexlab{b}})},\ \Eprint
  {http://arxiv.org/abs/2105.05872} {arXiv:2105.05872 [gr-qc]} \BibitemShut
  {NoStop}%
\bibitem [{\citenamefont {Zackay}\ \emph {et~al.}(2021)\citenamefont {Zackay},
  \citenamefont {Venumadhav}, \citenamefont {Roulet}, \citenamefont {Dai},\
  and\ \citenamefont {Zaldarriaga}}]{Zackay:2019kkv}%
  \BibitemOpen
  \bibfield  {author} {\bibinfo {author} {\bibfnamefont {B.}~\bibnamefont
  {Zackay}}, \bibinfo {author} {\bibfnamefont {T.}~\bibnamefont {Venumadhav}},
  \bibinfo {author} {\bibfnamefont {J.}~\bibnamefont {Roulet}}, \bibinfo
  {author} {\bibfnamefont {L.}~\bibnamefont {Dai}}, \ and\ \bibinfo {author}
  {\bibfnamefont {M.}~\bibnamefont {Zaldarriaga}},\ }\href {\doibase
  10.1103/PhysRevD.104.063034} {\bibfield  {journal} {\bibinfo  {journal}
  {Phys. Rev. D}\ }\textbf {\bibinfo {volume} {104}},\ \bibinfo {pages}
  {063034} (\bibinfo {year} {2021})},\ \Eprint
  {http://arxiv.org/abs/1908.05644} {arXiv:1908.05644 [astro-ph.IM]}
  \BibitemShut {NoStop}%
\bibitem [{\citenamefont {Nitz}\ \emph
  {et~al.}(2021{\natexlab{b}})\citenamefont {Nitz}, \citenamefont {Harry},
  \citenamefont {Willis}, \citenamefont {Biwer}, \citenamefont {Brown},
  \citenamefont {Pekowsky}, \citenamefont {Dal~Canton}, \citenamefont
  {Williamson}, \citenamefont {Dent}, \citenamefont {Capano}, \citenamefont
  {Massinger}, \citenamefont {Lenon}, \citenamefont {Nielsen},\ and\
  \citenamefont {Cabero}}]{pycbcgithub}%
  \BibitemOpen
  \bibfield  {author} {\bibinfo {author} {\bibfnamefont {A.~H.}\ \bibnamefont
  {Nitz}}, \bibinfo {author} {\bibfnamefont {I.~W.}\ \bibnamefont {Harry}},
  \bibinfo {author} {\bibfnamefont {J.~L.}\ \bibnamefont {Willis}}, \bibinfo
  {author} {\bibfnamefont {C.~M.}\ \bibnamefont {Biwer}}, \bibinfo {author}
  {\bibfnamefont {D.~A.}\ \bibnamefont {Brown}}, \bibinfo {author}
  {\bibfnamefont {L.~P.}\ \bibnamefont {Pekowsky}}, \bibinfo {author}
  {\bibfnamefont {T.}~\bibnamefont {Dal~Canton}}, \bibinfo {author}
  {\bibfnamefont {A.~R.}\ \bibnamefont {Williamson}}, \bibinfo {author}
  {\bibfnamefont {T.}~\bibnamefont {Dent}}, \bibinfo {author} {\bibfnamefont
  {C.~D.}\ \bibnamefont {Capano}}, \bibinfo {author} {\bibfnamefont {T.~J.}\
  \bibnamefont {Massinger}}, \bibinfo {author} {\bibfnamefont {A.~K.}\
  \bibnamefont {Lenon}}, \bibinfo {author} {\bibfnamefont {A.~B.}\ \bibnamefont
  {Nielsen}}, \ and\ \bibinfo {author} {\bibfnamefont {M.}~\bibnamefont
  {Cabero}},\ }\href@noop {} {\enquote {\bibinfo {title} {{PyCBC Software}},}\
  }\bibinfo {howpublished} {\url{https://github.com/gwastro/pycbc}} (\bibinfo
  {year} {2021}{\natexlab{b}})\BibitemShut {NoStop}%
\bibitem [{\citenamefont {Biwer}\ \emph {et~al.}(2019)\citenamefont {Biwer},
  \citenamefont {Capano}, \citenamefont {De}, \citenamefont {Cabero},
  \citenamefont {Brown}, \citenamefont {Nitz},\ and\ \citenamefont
  {Raymond}}]{Biwer:2018osg}%
  \BibitemOpen
  \bibfield  {author} {\bibinfo {author} {\bibfnamefont {C.~M.}\ \bibnamefont
  {Biwer}}, \bibinfo {author} {\bibfnamefont {C.~D.}\ \bibnamefont {Capano}},
  \bibinfo {author} {\bibfnamefont {S.}~\bibnamefont {De}}, \bibinfo {author}
  {\bibfnamefont {M.}~\bibnamefont {Cabero}}, \bibinfo {author} {\bibfnamefont
  {D.~A.}\ \bibnamefont {Brown}}, \bibinfo {author} {\bibfnamefont {A.~H.}\
  \bibnamefont {Nitz}}, \ and\ \bibinfo {author} {\bibfnamefont
  {V.}~\bibnamefont {Raymond}},\ }\href {\doibase 10.1088/1538-3873/aaef0b}
  {\bibfield  {journal} {\bibinfo  {journal} {Publ. Astron. Soc. Pac.}\
  }\textbf {\bibinfo {volume} {131}},\ \bibinfo {pages} {024503} (\bibinfo
  {year} {2019})},\ \Eprint {http://arxiv.org/abs/1807.10312} {arXiv:1807.10312
  [astro-ph.IM]} \BibitemShut {NoStop}%
\bibitem [{\citenamefont {Abbott}\ \emph
  {et~al.}(2021{\natexlab{b}})\citenamefont {Abbott} \emph
  {et~al.}}]{LIGOScientific:2019lzm}%
  \BibitemOpen
  \bibfield  {author} {\bibinfo {author} {\bibfnamefont {R.}~\bibnamefont
  {Abbott}} \emph {et~al.} (\bibinfo {collaboration} {LIGO Scientific,
  Virgo}),\ }\href {\doibase 10.1016/j.softx.2021.100658} {\bibfield  {journal}
  {\bibinfo  {journal} {SoftwareX}\ }\textbf {\bibinfo {volume} {13}},\
  \bibinfo {pages} {100658} (\bibinfo {year} {2021}{\natexlab{b}})},\ \Eprint
  {http://arxiv.org/abs/1912.11716} {arXiv:1912.11716 [gr-qc]} \BibitemShut
  {NoStop}%
\bibitem [{\citenamefont {{Speagle}}(2020)}]{2020MNRAS.493.3132S}%
  \BibitemOpen
  \bibfield  {author} {\bibinfo {author} {\bibfnamefont {J.~S.}\ \bibnamefont
  {{Speagle}}},\ }\href {\doibase 10.1093/mnras/staa278} {\bibfield  {journal}
  {\bibinfo  {journal} {"Mon. Not. Roy. Astron. Soc."}\ }\textbf {\bibinfo
  {volume} {493}},\ \bibinfo {pages} {3132} (\bibinfo {year} {2020})},\ \Eprint
  {http://arxiv.org/abs/1904.02180} {arXiv:1904.02180 [astro-ph.IM]}
  \BibitemShut {NoStop}%
\bibitem [{\citenamefont {Capano}(2021)}]{pykerr}%
  \BibitemOpen
  \bibfield  {author} {\bibinfo {author} {\bibfnamefont {C.~D.}\ \bibnamefont
  {Capano}},\ }\href@noop {} {\enquote {\bibinfo {title} {{pykerr}},}\
  }\bibinfo {howpublished} {\url{https://github.com/cdcapano/pykerr}} (\bibinfo
  {year} {2021})\BibitemShut {NoStop}%
\bibitem [{\citenamefont {Thrane}\ and\ \citenamefont
  {Talbot}(2019)}]{Thrane_2019}%
  \BibitemOpen
  \bibfield  {author} {\bibinfo {author} {\bibfnamefont {E.}~\bibnamefont
  {Thrane}}\ and\ \bibinfo {author} {\bibfnamefont {C.}~\bibnamefont
  {Talbot}},\ }\href {\doibase 10.1017/pasa.2019.2} {\bibfield  {journal}
  {\bibinfo  {journal} {Publications of the Astronomical Society of Australia}\
  }\textbf {\bibinfo {volume} {36}} (\bibinfo {year} {2019}),\
  10.1017/pasa.2019.2}\BibitemShut {NoStop}%
\end{thebibliography}%

\end{document}